# A generalized scattering theory in quantum mechanics

Huai-Yu Wang
Department of Physics, Tsinghua University, Beijing 100084 China
wanghuaiyu@mail.tsinghua.edu.cn

**Abstract**

In quantum mechanics textbooks, a single-particle scattering theory is introduced. In the present work, a generalized scattering theory is presented, which can be in principle applied to the scattering problems of arbitrary number of particle. In laboratory frame, a generalized Lippmann-Schwinger scattering equation is derived. We emphasized that the derivation is rigorous, even for treating infinitesimals. No manual operation such as analytical continuation is allowed. In the case that before scattering $N$ particles are plane waves and after the scattering they are new plane waves, the transition amplitude and transition probability are given and the generalized $S$ matrix is presented. It is proved that the transition probability from a set of plane waves to a new set of plane waves of the $N$ particles equal to that of the reciprocal process. The generalized theory is applied to the cases of one- and two-particle scattering as two examples. When applied to single-particle scattering problems, our generalized formalism degrades to that usually seen in the literature. When our generalized theory is applied to two-particle scattering problems, the formula of the transition probability of two-particle collision is given. It is shown that the transition probability of the scattering of two free particles is identical to that of the reciprocal process. This transition probability and the identity are needed in deriving Boltzmann transport equation in statistical mechanics. The case of identical particles is also discussed.

## 1. Introduction

The single-particle scattering theory in quantum mechanics (QM) is a mature theory with complete formalism. It is usually introduced in QM textbooks [1–6]. Trace to its source, Lippmann and Schwinger did the first work and they established the famous scattering equation named after them [7].

There are three motivations for the present work.

The first motivation is to generalize the single-particle scattering theory to many-particle one, so that one is able to deal with the scattering problems concerning arbitrary number of particle. In the literature, besides the single-particle scattering [7,8], the few-particle scattering problems have also been touched, e.g., [9–12]. Our aim is to give a uniform formalism applicable to arbitrary number of particle, as well as to identical particles. We will start from time-dependent Hamiltonians to obtain universal formulas. Time-independent Hamiltonians are special cases. When our theory is applied to the case of single-particle, the formalism of the scattering theory introduced in QM textbooks is naturally retrieved.

The second motivation is that the formalism should be derived rigorously, even for the occurrence of infinitesimals, without manual operation such as analytical continuation or manually inserting a factor into a formula, which is explained in Appendix A.

To specify this point, let us first recall how Lippmann and Schwinger obtain their time-independent scattering equation. We start from any time-dependent Hamiltonian. Suppose that the Hamiltonian $H$ of a system can be divided into two parts,

$$H = H_0 + H_1. \qquad (1.1)$$

Usually, the $H_0$ is chosen such that its eigenvalues and eigen functions are easily solved. For example, for an $N$-particle system, $H_0$ is chosen as the Hamiltonian of the $N$ noninteractive particle, so that its eigenvalues and eigenvectors are those of plane waves, and $H_1$ is chosen as the interaction between the particles. The Schrödinger equation corresponding to the $H$ and $H_0$ are respectively

$$(\mathrm{i}\hbar\frac{\partial}{\partial t} - H)\psi(t) = 0 \qquad (1.2)$$

and

$$(\mathrm{i}\hbar\frac{\partial}{\partial t} - H_0)\psi^{(0)}(t) = 0\,. \qquad (1.3)$$

Here, the spatial arguments of the wave functions are not explicitly written. If the Hamiltonian (1.1) is time-independent, a time factor in the exponential form can be separated from the wave functions as follows.

$$\psi(t) = \varphi \mathrm{e}^{-\mathrm{i}Et}\,.\ \ \psi^{(0)}(t) = \varphi^{(0)}\mathrm{e}^{-\mathrm{i}E_0 t}\,. \qquad (1.4)$$

Lippmann and Schwinger inserted an exponential factor $\mathrm{e}^{-\varepsilon|t|}$, where $\varepsilon \to 0^+$, into the integral expression of the solution $\psi(t)$ such that they obtained a time-independent equation [7,8,10,11]:

$$\varphi = \varphi^{(0)} + \frac{1}{E - H_0 + \mathrm{i}\varepsilon} H_1\varphi, \varepsilon \to 0^+\,. \qquad (1.5)$$

Hereafter, the $\mathrm{i}\varepsilon, \varepsilon \to 0^+$ is shortened as $\mathrm{i}0^+$. Since Eq. (1.5) appeared, it has often been employed as the starting point of studying scattering problems [8–34]. Another way of obtaining (1.5) is to make the $E$ in Green's function $G^{(0)}(E) = \frac{1}{E - H_0}$ deviate from the real axis. We think that inserting the factor $\mathrm{e}^{-\varepsilon|t|}$ and making analytical continuation belong to manual operation but not rigorous derivation. Analytical continuation may not be unique, e.g., Eqs. (6) and (7) in Ref. [35].

The aim of introducing the infinitely small imaginary part $\mathrm{i}0^+$ in the denominator of (1.5) was to reflect the effect of time delay. One "used an infinitesimal imaginary energy part $\pm\mathrm{i}\varepsilon$ to obtain, respectively, the incoming and outgoing solutions of the Lippmann–Schwinger equations, and distinguished between "states at time $t' < t_0$ =time defined by preparation" and "states characteristic of the experiment," observed at $t'' > t_0$." [36] This infinitesimal imaginary part was thought as a boundary condition. When (1.5) is written in the form of $|E^{\pm}\rangle = |E\rangle + \frac{1}{E - H_0 \pm \mathrm{i}\varepsilon} H_1 |E^{\pm}\rangle$, (see Eq. (2.3±) in [36]), then, $|E^{\pm}\rangle$ "fulfill different boundary conditions expressed by +i0 and –i0." [36]

Since the $\pm\mathrm{i}0^+$ in the denominator in the Lippmann-Schwinger equation respectively show the significance of time delay and advance, they ought to be derived naturally, but not added through a manual operation of artificially inserting a factor $\mathrm{e}^{-\varepsilon|t|}$ or of making analytical continuation. In a previous work by the author [35], it was pointed out that the time step function $\theta(t - t_0)$ could clearly express the meaning of time delay. That this factor was contained in a wave function meant that we were considering an evolution of a system at any time $t > t_0$ where $t_0$ is the initial instant. The Fourier transformation of the $\theta(t - t_0)$ is

$$\theta(t - t_0) = \frac{\mathrm{i}}{2\pi} \int_{-\infty}^{\infty} \frac{\mathrm{e}^{-\mathrm{i}\varepsilon(t - t_0)}}{\varepsilon + \mathrm{i}0^+} \mathrm{d}\varepsilon . \quad (1.6)$$

There was a detailed discussion of how to achieve this transformation [35]. Its derivative is [35,37,38]

$$\frac{\partial}{\partial t}\theta(t-t_0)=\delta(t-t_0)-0^+\theta(t-t_0). \tag{1.7}$$

Usually, the infinitesimal is ignored in the literature. We stress that it should not be ignored, because it is of the physical meaning of time delay. The Dirac delta function $\delta(t-t_0)$ itself does not reflect the meaning of time delay, but $\theta(t-t_0)$ does. Since in Eq. (1.7) the left hand side is of the meaning of time delay, the right hand side should also be. The infinitesimal in (1.7) comes from the expression (1.6), which correctly embodies time delay, and it is also the source of the infinitesimal imaginary part in the denominator of (1.5). It seemed that there was only one researcher [39] who was aware of that the infinitesimal in (1.5) was from the Fourier transformation (1.6).

The third motivation is to prove an identity needed in statistical mechanics.

It is well known that derivation of Boltzmann transport equation involves an important physical quantity $w(\boldsymbol{p}_1,\boldsymbol{p}_2,\boldsymbol{p}_1',\boldsymbol{p}_2';t)$ that is the transition probability of two-particle collision. Suppose that there are two free particles with momenta $\boldsymbol{p}_1$ and $\boldsymbol{p}_2$, respectively. They collide with each other, and then transit to new free states with momenta $\boldsymbol{p}_1'$ and $\boldsymbol{p}_2'$, respectively. The quantity $w(\boldsymbol{p}_1,\boldsymbol{p}_2,\boldsymbol{p}_1',\boldsymbol{p}_2';t)$ means the transition probability of such a collision process. In establishing Boltzmann transport equation, it was assumed that the following identity stood:

$$w(\boldsymbol{p}_1,\boldsymbol{p}_2,\boldsymbol{p}_1',\boldsymbol{p}_2';t)=w(\boldsymbol{p}_1',\boldsymbol{p}_2',\boldsymbol{p}_1,\boldsymbol{p}_2;t). \tag{1.8}$$

"The function $w$ can in principle be determined only by solving the mechanical problem of collision of particles interacting according to some given law. However, certain properties of this function can be elucidated from general arguments." After these words, it was argued that the assumption of (1.8) was reasonable [40]. In [41], $w(\boldsymbol{p}_1,\boldsymbol{p}_2,\boldsymbol{p}_1',\boldsymbol{p}_2';t)$ and $w(\boldsymbol{p}_1',\boldsymbol{p}_2',\boldsymbol{p}_1,\boldsymbol{p}_2;t)$ were respectively denoted as $a$ and $a'$, and assumed $a'=a$, (see Eqs. (D1.20), (D1.03), and (D1.05) in [41]). In [42], the same assumption was made: "In general, the symmetry does not necessarily hold. However, let us assume this for simplicity." In [43,44], Eq. (1.8) was illustrated by graphical manner without rigorous mathematical derivation.

In short, Eq. (1.8) is assumed to be true, but no proof has been given in QM. The reason is that when discussing the collision between two particles in the literature, usually, the mass-center frame, but not laboratory frame, is adopted such that the two-particle collision problem is reduced to be one-particle scattering [2,3,6]. The formalism for one-particle scattering is unable to give the expression of $w(\boldsymbol{p}_1,\boldsymbol{p}_2,\boldsymbol{p}_1',\boldsymbol{p}_2';t)$. Without this expression, Eq. (1.8) cannot be proved.

The generalized scattering theory is in principle able to deal with the problems of scattering of an arbitrary number of particle in laboratory frame. In the case of two-particle collision, the expression of $w(\boldsymbol{p}_1, \boldsymbol{p}_2, \boldsymbol{p}'_1, \boldsymbol{p}'_2; t)$ is put down, and subsequently, (1.8) is proved.

This paper is arranged as follows. In Section 2, the generalized scattering theory is presented. A general time-dependent Lippmann-Schwinger equation (see Eqs. (2.36) and (2.37) below) and time-independent equation (1.5) (also see Eq. (2.41) below, which is obtained under the condition that the Hamiltonian is time-independent) are derived. The generalized *S* matrix is put forth, and its unitary is proved. In Section 3, the theory is applied to the case of single-particle scattering. The discrepancy between two scattering pictures appeared in textbooks (see Figs. 1 and 2 below) is conceptually clarified. In Section 4, the theory is applied to the case of two-particle collision, and Eq. (1.8) is proved. Our conclusion is in Section 5.

In this work, we merely discuss elastic scattering. We do not consider the inner structures of colliding particles, and do not consider inelastic collision. The two words scattering and collision are regarded as synonym. Laboratory frame is employed. We choose laboratory frame because all experiments are done in laboratory frame. In textbooks, when treating the problem of two-particle collision, usually, the mass-center frame is adopted as mentioned above. However, the results have to be transformed to be those in the laboratory frame so as to compare them with experimental ones. Moreover, for the problems of collisions of more than two particles, the mass-center frame is very difficult to use.

## 2. The generalized scattering theory

Suppose that a system contains *N* particles and its Hamiltonian is *H*. The set of all the spatial coordinates in this system is denoted by an italic letter *R*. The Schrödinger equation is

$$(\mathrm{i}\hbar\frac{\partial}{\partial t} - H)\psi(R,t) = 0\,. \tag{2.1}$$

In the following formulas, the spatial arguments may be dropped and only the time argument is explicitly shown.

Equation (2.1) is a differential equation with the first derivative of time. In order to solve such an equation for a concrete system, an initial condition is required. The initial time is denoted by $t_0$. The initial condition sets the values of the solutions at time $t_0$. The solved solutions apply for the time $t > t_0$, i.e., the time-delay or time-retarded solutions, which was mentioned previously [35]. Hereafter, we use a superscript regular letter R to represent the meaning of time delay. For examples, $\psi^{\mathrm{R}}$ and $G^{\mathrm{R}}$ respectively mean retarded wave function and retarded Green's function.

When discussing a time-delay solution, the solution necessarily contains a factor of time step function $\theta(t-t_0)$. For example, Lippmann and Schwinger mentioned this in their work [7]. Some authors [4,45–48], usually in discussing scattering problems, have noticed this: a factor of time step function $\theta(t-t_0)$ ought to be attached, although they did not address this through derivation of formulas. Since the factor $\theta(t-t_0)$ ought to appear in the solutions, it should also appear in the differential equation. This is because the solutions are solved from the differential equation.

A retarded wave function obeys the following equation [35]:

$$(\mathrm{i}\hbar\frac{\partial}{\partial t}-H)\psi^{\mathrm{R}}(t)=\mathrm{i}\hbar\frac{\partial\theta(t-t_0)}{\partial t}\psi(t_0), \tag{2.2}$$

where $\psi(t_0)$ is the value of the retarded function $\psi^{\mathrm{R}}$ at instant $t_0$. The important significance of Eq. (2.2) is that the initial condition is explicitly included in the differential equation itself. Usually, a differential equation and its initial conditions are written separately, the consequence of which is that the role of the initial conditions may be looked down upon or even ignored.

It should be noticed that in (2.2) the factor $\frac{\partial\theta(t-t_0)}{\partial t}$ is not simplified to be $\delta(t-t_0)$, because this factor should imply the meaning of the time retard. We have stressed that the result of $\frac{\partial\theta(t-t_0)}{\partial t}$ contained an infinitesimal, (see Eq. (1.7)).

The solution $\psi^{\mathrm{R}}(t)$ of (2.2) should contain the information of time retard, so that it is written in the following form:

$$\psi^{\mathrm{R}}(R,t)=\theta(t-t_0)\psi(R,t). \tag{2.3}$$

When Eq. (2.3) is substituted into (2.2), we obtain the Schrödinger equations (2.1) and initial condition

$$\psi(R,t_0)=\psi(R,t)|_{t=t_0}. \tag{2.4}$$

That is to say, Eq, (2.2) contains the Schrödinger equation (2.1) and the initial condition (2.4) simultaneously. Please note that in (2.4), *t* approaches $t_0$ at the side of $t>t_0$.

The retarded Green's function of the *H* system obeys the following equation [35]:

$$(\mathrm{i}\hbar\frac{\partial}{\partial t}-H)G^{\mathrm{R}}(R,R';t,t_0)=\mathrm{i}\hbar\delta(R-R')\frac{\partial}{\partial t}\theta(t-t_0). \tag{2.5}$$

Similar to the retarded wave function, the $G^{\mathrm{R}}(R,R';t,t_0)$ should also contain a time step function.

$$G^{\mathrm{R}}(t,t_0)=\theta(t-t_0)G(t,t_0). \tag{2.6}$$

Substitution of (2.6) into (2.5) leads to the equation satisfied by the $G$

$$(\mathrm{i}\hbar\frac{\partial}{\partial t}-H)G(R,R';t,t')=0. \tag{2.7}$$

and the initial condition

$$G(R,R';t,t_0)|_{t=t_0}=\delta(R-R'). \tag{2.8}$$

It is seen that Eq. (2.5) contains Eq. (2.7) and the initial condition (2.8) simultaneously. We stress that it is $G^{\mathrm{R}}$ that is of the meaning of time delay. The $G$ reflects neither time delay nor time advance. So, it can be called an auxiliary function [35]. We simply call the $G$ as Green's function, consistent with habit.

If the $G^{\mathrm{R}}$ is worked out, it can be employed to find the retarded solution of (2.2) through the following procedure. An inverse operator $(\mathrm{i}\hbar\frac{\partial}{\partial t}-H)^{-1}$ is acted on both sides of (2.2). The result is

$$\begin{aligned}\psi^{\mathrm{R}}(R,t)&=(\mathrm{i}\hbar\frac{\partial}{\partial t}-H)^{-1}\mathrm{i}\hbar\frac{\partial\theta(t-t_0)}{\partial t}\psi(R,t_0)\\&=\int\mathrm{d}R'(\mathrm{i}\hbar\frac{\partial}{\partial t}-H)^{-1}\mathrm{i}\hbar\frac{\partial\theta(t-t_0)}{\partial t}\delta(R-R')\psi(R',t_0)\\&=\int\mathrm{d}R'G^{\mathrm{R}}(R,R';t,t_0)\psi(R',t_0).\end{aligned} \tag{2.9a}$$

Literally, this formula demonstrates that the $G^{\mathrm{R}}(R,R';t,t_0)$ plays such a role that it helps to propagate one state to another through all possible intermediate paths. Physically, the $\psi^{\mathrm{R}}(R,t)$ is a state of the Hamiltonian $H$, while $\psi(R,t_0)$ is the initial condition. The significance of (2.9a) is that when the state $\psi(R,t_0)$ is acted by the retarded Green's function of the system $H$, it evolves into a state of the $H$. Therefore, $\psi(R,t_0)$ can be called an initial state, and $\psi^{\mathrm{R}}(R,t)$ final state at time $t$.

We substitute Eqs. (2.3) and (2.6) into (2.9a). When $t>t_0$, $\theta(t-t_0)$ takes 1. Consequently, (2.9a) is simplified to be

$$\psi(R,t)=\int\mathrm{d}R'G(R,R';t,t_0)\psi(R',t_0),\ t>t_0. \tag{2.9b}$$

This is a fundamental formula in path integral [49].

In this work, we merely discuss retarded solutions. So, hereafter, if in a formula the factor $\theta(t-t_0)$ or the superscript R does not appear, there is an acquiescence that

$$t > t_0 . \tag{2.10}$$

We point out that it is Eq. (2.9a) that is rigorous. Equation (2.9b) is actually a simplified form of (2.9a) as (2.10) is satisfied.

Often, it is hard to exactly solve the wave functions and Green's functions of a Hamiltonian *H*. A commonly used method is dividing a Hamiltonian *H* into two parts,

$$H = H_0 + H_1 , \tag{2.11}$$

where the $H_0$ is chosen such that its eigen wave functions $\psi^{(0)}(R,t)$ and corresponding Green's function $G^{(0)}$ can be easily solved (see Eqs. (2.12) and (2.15) below), and $H_1$ is the remaining part of the *H*. Here, a superscript (0) marks the quantities belonging to $H_0$ system.

The above formulas (2.1)-(2.9) are valid to any Hamiltonian *H*, and valid to $H_0$ too. We simply make the following substitutions in (2.1)-(2.9): *H* is replaced by $H_0$, $\psi(t)$ by $\psi^{(0)}(t)$, $\psi^{\mathrm{R}}(R,t)$ by $\psi^{\mathrm{R}(0)}(R,t)$, $G(t)$ by $G^{(0)}(t)$, $G^{\mathrm{R}}(t)$ by $G^{\mathrm{R}(0)}(t)$, and $\psi(t_0)$ by $\psi^{(0)}(t_0)$. In this way, the following formulas concerning $H_0$ are acquired.

The retarded wave function $\psi^{\mathrm{R}(0)}(t)$ obeys the equation

$$(\mathrm{i}\hbar\frac{\partial}{\partial t} - H_0)\psi^{\mathrm{R}(0)}(t) = \mathrm{i}\hbar\frac{\partial\theta(t-t_0)}{\partial t}\psi^{(0)}(t_0) . \tag{2.12}$$

Its solutions $\psi^{\mathrm{R}(0)}(t)$ are of the form

$$\psi^{\mathrm{R}(0)}(t) = \theta(t-t_0)\psi^{(0)}(t) . \tag{2.13}$$

It follows that Eq. (2.12), when (2.13) is used, contains both the Schrödinger equation

$$(\mathrm{i}\hbar\frac{\partial}{\partial t} - H_0)\psi^{(0)}(t) = 0 . \tag{2.14a}$$

and an initial condition

$$\psi^{(0)}(t_0) = \psi^{(0)}(t)\,|_{t=t_0} . \tag{2.14b}$$

Correspondingly, the retarded Green's function of $H_0$ satisfies the equation

$$(\mathrm{i}\hbar\frac{\partial}{\partial t}-H_0)G^{\mathrm{R}(0)}(R,R';t,t_0)=\mathrm{i}\hbar\delta(R-R')\frac{\partial}{\partial t}\theta(t-t_0)\,. \tag{2.15}$$

Its solution should also have a time step function factor.

$$G^{\mathrm{R}(0)}=\theta(t-t_0)G^{(0)}\,. \tag{2.16}$$

After (2.16) is substituted into (2.15), we obtain the equation satisfied by $G^{(0)}$,

$$(\mathrm{i}\hbar\frac{\partial}{\partial t}-H_0)G^{(0)}(R,R';t,t_0)=0\,, \tag{2.17}$$

and an initial condition

$$G^0(R,R';t,t_0)\,|_{t=t_0}=\delta(R-R')\,. \tag{2.18}$$

The retarded wave function can be expressed by the retarded Green's function. From (2.9a), it is known that

$$\psi^{\mathrm{R}(0)}(R,t)=\int \mathrm{d}R'G^{\mathrm{R}(0)}(R,R';t,t_0)\psi^{(0)}(R',t_0)\,. \tag{2.19a}$$

When (2.10) is defaulted, the time step functions can be dropped and we have

$$\psi^{(0)}(R,t)=\int \mathrm{d}R'G^{(0)}(R,R';t,t_0)\psi^{(0)}(R',t_0)\,. \tag{2.19b}$$

Equation (2.19) has the same physical significance as (2.9). The former is just for the $H_0$ system. Both equations can be iterated repeatedly. The time step functions in these formulas guarantee the semigroup evolution in time order [36].

It is a customary choosing a time-independent $H_0$ the eigenenergies $\{E_n\}$ and corresponding eigen wave functions $\{\varphi_n^{(0)}\}$ of which are easily solved. Note that we are considering a system containing an arbitrary number of particle, so that here the subscript *n* actually represents a set of quantum numbers. The stationary equation is

$$H_0\varphi_n^{(0)}(R)=E_n\varphi_n^{(0)}(R)\,. \tag{2.20}$$

Then, the stationary wave function is

$$\psi_n^{(0)}(R,t)=\varphi_n^{(0)}(R)\mathrm{e}^{-\mathrm{i}E_nt/\hbar}\,. \tag{2.21}$$

All the eigen wave functions constitute a complete set.

$$\sum\nolimits_n\varphi_n^{(0)}(R)\varphi_n^{(0)*}(R')=\delta(R-R')\,. \tag{2.22}$$

The $G^{(0)}(R,R';t,t_0)$ can be expressed by the complete set.

$$G^{(0)}(R,R';t,t_0)=\sum\nolimits_n\psi_n^{(0)}(R,t)\psi_n^{(0)*}(R',t_0)=\mathrm{e}^{-\mathrm{i}H_0(t-t_0)/\hbar}\delta(R-R')\,, \tag{2.23}$$

where use of (2.20)-( 2.22) is made. The last form is the action of an operator on the

Dirac delta function. It follows from (2.23) that

$$G^{(0)*}(R',R;t_0,t)=G^{(0)}(R,R';t,t_0)\,. \qquad (2.24)$$

If a function $\psi^{(0)*}(R,t)$ can be expanded as

$$\psi^{(0)*}(R,t)=\sum_m a_m(t)\varphi_m^{(0)*}(R)\mathrm{e}^{\mathrm{i}E_m t/\hbar}\,, \qquad (2.25)$$

then, it can be proved by use of (21)-(23) that

$$\psi^{(0)*}(R,t)=\int \mathrm{d}R'\psi^{(0)*}(R',t_0)G^{(0)}(R',R;t_0,t)\,. \qquad (2.26a)$$

The physical significance of (2.26a) is similar to that of (2.9b).

Similarly, suppose that a Hamiltonian $H$ is time-independent and its eigen spectrum is easily solved. If a function $\psi^*(R,t)$ can be expanded by the complete set of the eigen functions of the $H$, just as (2.25), then we have

$$\psi^*(R,t)=\int \mathrm{d}R'\psi^*(R',t_0)G(R',R;t_0,t)\,, \qquad (2.26b)$$

which corresponds to (2.9a). It is noted that (2.9) is valid for any Hamiltonian, while (2.26) is proved only for time-independent Hamiltonian.

Once the quantities of $H_0$ system are gained, they can be utilized, together with the interaction $H_1$, to find the quantities of the $H$ system. For instance, the $G^{\mathrm{R}(0)}$ and $H_1$ can be used to express the $G^{\mathrm{R}}$. We substitute (2.11) into (2.5) and obtain

$$(\mathrm{i}\hbar\frac{\partial}{\partial t}-H_0)G^{\mathrm{R}}=\mathrm{i}\hbar\delta(R-R')\frac{\partial\theta(t-t_0)}{\partial t}+H_1G^{\mathrm{R}}\,. \qquad (2.27)$$

An inverse operator $(\mathrm{i}\hbar\frac{\partial}{\partial t}-H)^{-1}$ is acted on both sides of (2.27), and then, the insertion of $\int \mathrm{d}R'\delta(R-R')$ and $\int_{-\infty}^{\infty}\mathrm{d}t_1\delta(t-t_1)$ results in

$$\begin{aligned}G^{\mathrm{R}}&=(\mathrm{i}\hbar\frac{\partial}{\partial t}-H_0)^{-1}\mathrm{i}\hbar\delta\frac{\partial\theta}{\partial t}+(\mathrm{i}\hbar\frac{\partial}{\partial t}-H_0)^{-1}H_1G^{\mathrm{R}}\\&=G^{\mathrm{R}(0)}+\frac{1}{\mathrm{i}\hbar}\int_{-\infty}^{\infty}\mathrm{d}t_1\int \mathrm{d}R_1G^{\mathrm{R}(0)}(R,R_1;t,t_1)H_1G^{\mathrm{R}}(R_1,R';t_1,t_0).\end{aligned} \qquad (2.28)$$

This equation can be iterated repeatedly. As an abbreviation, we drop the integrals in (2.28) to simplify it to be the products of the functions.

$$G^{\mathrm{R}}=G^{\mathrm{R}(0)}+G^{\mathrm{R}(0)}H_1G^{\mathrm{R}}\,. \qquad (2.29)$$

The result of the iteration is

$$G^{\mathrm{R}}=G^{\mathrm{R}(0)}+G^{\mathrm{R}(0)}TG^{\mathrm{R}(0)}\,, \qquad (2.30)$$

where

$$T = H_1 + H_1 G^{\mathrm{R}(0)} H_1 + H_1 G^{\mathrm{R}(0)} H_1 G^{\mathrm{R}(0)} H_1 + \cdots = H_1 + H_1 G^{\mathrm{R}(0)} T \,. \tag{2.31}$$

The $T$ was called $t$ matrix, a short form of transition matrix [6]. Strictly speaking, $T$ should also have a superscript R. Nevertheless, dropping the superscript does not affect the discussion below. Please note that all the Green's functions in (2.29)-(2.31) are retarded ones, each carrying a time step function. Using Eqs. (2.6) and (2.16), the superscripts R in Eqs. (2.29)-(2.31) can be removed, and the three formulas remain valid.

Furthermore, if the $H_1$ is real, i.e.,

$$H_1^*(R) = H_1(R) \,, \tag{2.32}$$

then, the $T$ defined by (2.31) is of the property

$$T^*(R) = T(R) \,. \tag{2.33}$$

We suppose that at time $t_0$, the wave function $\psi^{(0)}(t_0)$ is just a state of $H_0$.

$$\psi(t_0) = \psi^{(0)}(t_0) \,. \tag{2.34}$$

Then, Eq. (2.9) is written as

$$\psi^{\mathrm{R}}(R,t) = \int \mathrm{d}R' G^{\mathrm{R}}(R,R';t,t_0)\psi^{(0)}(R',t_0) \,. \tag{2.35}$$

This equation has the same physical meaning as that of (2.9), with the initial state $\psi^{(0)}(R,t)$ just belonging to the $H_0$. The final state belongs to the $H$.

Substituting (2.28) into (2.35) and making use of (2.19) and (2.26), we obtain

$$\begin{aligned} \psi^{\mathrm{R}}(R,t) &= \psi^{\mathrm{R}(0)}(R,t) \\ &+ \frac{1}{\mathrm{i}\hbar}\int_{-\infty}^{\infty} \mathrm{d}t_1 \int \mathrm{d}R' \int \mathrm{d}R_1 G^{\mathrm{R}(0)}(R,R_1;t,t_1) H_1 G^{\mathrm{R}}(R_1,R';t_1,t_0)\psi^{(0)}(R',t_0) \\ &= \psi^{\mathrm{R}(0)}(R,t) + \frac{1}{\mathrm{i}\hbar}\int_{-\infty}^{\infty} \mathrm{d}t_1 \int \mathrm{d}R_1 G^{\mathrm{R}(0)}(R,R_1;t,t_1) H_1 \psi^{\mathrm{R}}(R_1,t_1). \end{aligned} \tag{2.36}$$

Repeated iteration of this equation, with the help of (2.30) and (2.31), leads to

$$\psi^{\mathrm{R}}(R,t) = \psi^{\mathrm{R}(0)}(R,t) + \frac{1}{\mathrm{i}\hbar}\int_{-\infty}^{\infty} \mathrm{d}t_1 \int \mathrm{d}R_1 G^{\mathrm{R}(0)}(R,R_1;t,t_1) T \psi^{\mathrm{R}(0)}(R_1,t_1) \,. \tag{2.37}$$

Equations (2.36) and (2.37) are formally the same as Lippmann-Schwinger equation, so that can be called generalized Lippmann-Schwinger equation, which can be used to systems composed of any number of particle.

We explicitly write the step functions in (2.37).

$$\begin{aligned} \psi(R,t)\theta(t-t_0) &= \psi^{(0)}(R,t)\theta(t-t_0) \\ &+ \frac{1}{\mathrm{i}\hbar}\int_{-\infty}^{\infty} \mathrm{d}t_1 \int \mathrm{d}R_1 \theta(t-t_1) G^{(0)}(R,R_1;t,t_1) H_1(R_1)\psi(R_1,t_1)\theta(t_1-t_0). \end{aligned} \tag{2.38}$$

Up to now, the derivation is rigorous, and Hamiltonian $H$ can be an arbitrary one.

If the Hamiltonian is time-independent, the formulas above can be simplified somewhat. The time factor of a wave function can be separated, as in the form of Eq. (1.4). Then, (2.37) becomes

$$\begin{aligned}&\psi(R)\theta(t-t_0)\\&=\varphi^{(0)}(R)\theta(t-t_0)+\frac{1}{\mathrm{i}\hbar}\int_{-\infty}^{\infty}\mathrm{d}t_1\theta(t-t_1)\mathrm{e}^{-\mathrm{i}(H_0-E)(t-t_1)}H_1(R)\psi(R)\theta(t_1-t_0).\end{aligned} \tag{2.39}$$

This equation is exact, as long as Hamiltonian is time-independent.

We are considering the solution at the time satisfying (2.10). For $t_1 > t_0$, we let in (2.39)

$$\theta(t-t_0)=1 \text{ and } \theta(t_1-t_0)=1. \tag{2.40}$$

That is to say, all the time-retard factors attached to the wave functions can be omitted, but that attached to the Green's function should not. Thus, the step function $\theta(t-t_1)$ in (2.39) is retained, and it is Fourier transformed by use of (1.6). Consequently, we have

$$\varphi(R)=\varphi^{(0)}(R)+\frac{1}{E-H_0+\mathrm{i}0^+}H_1\varphi(R). \tag{2.41}$$

This is generalized Lippmann-Schwinger equation for time-independent systems. As a comparison, Eq. (2.39) applies to any time-dependent Hamiltonian. By the way, the formulas in Appendix B in [12] were derived from (2.41). It is easily shown that from (2.39), the same formulas can be derived.

Here, we stress the differences between Eq. (2.41) and the original Lippmann-Schwinger equation, although they are of the same form. One is that (2.41) is not simply a single-particle equation. The Hamiltonian H (2.11) can contain arbitrary number of particles. Another one is that the infinitesimal in (2.41) is not from manual analytic continuation, but from Eq. (2.15) and its solution (2.16) where time retard has been embodied.

Because of the rigorous derivation above and implicit geralized physical meaning, we are able to do the following work.

Now, we further assume that the initial state $\psi^{(0)}$ in (2.37) is just one of the $H_0$'s eigenstates, i.e., one satisfying (2.21), denoted by

$$\psi^{(0)}=\psi_n^{(0)}. \tag{2.42}$$

The corresponding final state is denoted as $\psi_n$. Then, (2.37) becomes

$$\psi_n^{\mathrm{R}}(R,t)=\psi_n^{\mathrm{R}(0)}(R,t)+\frac{1}{\mathrm{i}\hbar}\int_{-\infty}^{\infty}\mathrm{d}t_1\int\mathrm{d}R_1G^{\mathrm{R}(0)}(R,R_1;t,t_1)T\psi_n^{\mathrm{R}(0)}(R_1,t_1). \tag{2.43}$$

The physical meaning of this equation is more explicit. The initial state $\psi_n^{(0)}$ is an eigen wave function of the $H_0$. When acted by the Hamiltonian *H*, it evolves into a final state $\psi_n$ that belongs to the *H*.

The eigen wave functions $\{\psi_n^{(0)}\}$ of the $H_0$ constitute a complete set. Any function can be expanded by this set, and so can the $\psi_n$ in (2.43). The projection of the $\psi_n$ onto a $\psi_m^{(0)}$ is the transition amplitude from the $\psi_n^{(0)}$ to $\psi_m^{(0)}$, denoted by $C_{mn}$.

$$\begin{aligned}C_{mn}(t) &= \int \mathrm{d}R\psi_m^{\mathrm{R}(0)*}(R,t)\psi_n^{\mathrm{R}}(R,t) \\ &= \int \mathrm{d}R\psi_m^{(0)*}(R,t)\psi_n^{\mathrm{R}(0)}(R,t) + \frac{1}{\mathrm{i}\hbar}\int_{-\infty}^{\infty}\mathrm{d}t_1\int \mathrm{d}R\psi_m^{\mathrm{R}(0)*}(R_1,t)T\psi_n^{\mathrm{R}(0)}(R,t_1).\end{aligned} \tag{2.44}$$

With the help of the second expression of the $G^{\mathrm{R}(0)}$ in (2.23), any order term of (2.44) can be expressed uniformly, and they are presented in Appendix B.

The transition probability from the $\psi_n^{(0)}$ to $\psi_m^{(0)}$ is

$$w_{mn}(t) = |C_{mn}(t)|^2. \tag{2.45}$$

On the other hand, we can let the initial state be

$$\psi^{(0)} = \psi_m^{(0)}. \tag{2.46}$$

Then, the corresponding final state is obtained by replacing *n* in (2.43) by *m*. This final state can also be expanded by the compete set $\{\psi_n^{(0)}\}$, and its projection onto the $\psi_n^{(0)}$ is denoted by $C_{nm}$, which can be put down simply by exchanging the subscripts *n* and *m* in (2.44). The corresponding transition probability is

$$w_{nm}(t) = |C_{nm}(t)|^2. \tag{2.47}$$

The transitions from the $\psi_n^{(0)}$ to $\psi_m^{(0)}$ and from the $\psi_m^{(0)}$ to $\psi_n^{(0)}$ are a pair of reciprocal or inverse transition processes, with $C_{mn}$ and $C_{nm}$ being corresponding transition amplitudes.

Furthermore, assume that the $H_1$ satisfies (2.32). Then, (2.33) is also satisfied. In this case, (2.44) is simplified to be

$$C_{mn}(t)=\int \mathrm{d}R\psi_m^{\mathrm{R}(0)*}(R,t)\psi_n^{\mathrm{R}(0)}(R,t)$$
$$+\frac{1}{\mathrm{i}\hbar}\int_{-\infty}^{\infty}\mathrm{d}t_1\int \mathrm{d}R_1\psi_m^{\mathrm{R}(0)*}(R_1,t_1)T\psi_n^{\mathrm{R}(0)}(R_1,t_1). \tag{2.48}$$

It follows that

$$C_{nm}(t)=C_{mn}^*(t). \tag{2.49}$$

Subsequently,

$$w_{nm}(t)=w_{mn}(t). \tag{2.50}$$

Suppose that there are $N$ free particles. After they collide with each other, they become free again. This is the transition from a set of plane waves to a new set of plane waves. Equation (2.50) reveals that the transition probability of this process equals to that of the reciprocal process.

Because the Hamiltonian $H_0(R)$ is time independent, its eigen functions are of the form (2.21). We let $\mathrm{e}^{-\mathrm{i}E_n t/\hbar}|\varphi_n^{(0)}\rangle=\mathrm{e}^{-\mathrm{i}H_0 t/\hbar}|\varphi_n^{(0)}\rangle$. Equation (2.48) is recast to be

$$C_{mn}(t)=\langle\varphi_m^{(0)}|\varphi_n^{(0)}\rangle+\langle\varphi_m^{(0)}|\int_{t_0}^{t}\mathrm{d}t_1\mathrm{e}^{\mathrm{i}H_0t_1/\hbar}T\mathrm{e}^{-\mathrm{i}H_0t_1/\hbar}|\varphi_n^{(0)}\rangle$$
$$=\langle\varphi_m^{(0)}|U(t,t_0)|\varphi_n^{(0)}\rangle. \tag{2.51}$$

Please note that because of the time step function, the upper and lower limits of the integral are actually $\int_{t_0}^{t}\mathrm{d}t_1$. In (2.51), an operator is defined:

$$U(t,t_0)=1+\int_{t_0}^{t}\mathrm{d}t_1\mathrm{e}^{\mathrm{i}H_0t_1/\hbar}T\mathrm{e}^{-\mathrm{i}H_0t_1/\hbar}. \tag{2.52}$$

It can be called generalized time evolution operator. Equation (2.49) reflects the unitary of the $U$ matrix.

When collisions between particles happen, usually, the time of collision is much less than the time that a free particle takes going through its mean free path. Therefore, in evaluation, the upper and lower limits of the integral in (2.52) can be extended to positive and negative infinities.

$$S=\lim_{\substack{t\to+\infty\\ t_0\to-\infty}}U(t,t_0)=1+\int_{-\infty}^{\infty}\mathrm{d}t_1\mathrm{e}^{\mathrm{i}H_0t_1/\hbar}T\mathrm{e}^{-\mathrm{i}H_0t_1/\hbar}. \tag{2.53}$$

This is generalized $S$ matrix. It is unitary.

$$S^+S=1. \tag{2.54}$$

The second term in (2.53) is denoted by $T_1$.

$$T_1=\int_{-\infty}^{\infty}\mathrm{d}t_1\mathrm{e}^{\mathrm{i}H_0t_1/\hbar}T\mathrm{e}^{-\mathrm{i}H_0t_1/\hbar}. \tag{2.55}$$

Then, it follows from (2.54) that

$$T_1^+ T = -(T + T_1^+)\,. \tag{2.56}$$

Now, we consider the case $m \neq n$ in (2.51). The transition probability is

$$w_{mn} = |\langle \varphi_m^{(0)} | T_1 | \varphi_n^{(0)} \rangle|^2\,. \tag{2.57}$$

If $H_1$ is time independent, we have

$$\begin{aligned} \langle \varphi_m^{(0)} | T_1 | \varphi_n^{(0)} \rangle &= \langle \varphi_m^{(0)} | \int_{-\infty}^{\infty} \mathrm{d}t \mathrm{e}^{\mathrm{i}H_0 t/\hbar} T \mathrm{e}^{-\mathrm{i}H_0 t/\hbar} | \varphi_n^{(0)} \rangle \\ &= \int_{-\infty}^{\infty} \mathrm{d}t \mathrm{e}^{-\mathrm{i}(E_n - E_m)t/\hbar} \langle \varphi_m^{(0)} | T | \varphi_n^{(0)} \rangle = 2\pi \delta(E_n - E_m) \langle \varphi_m^{(0)} | T | \varphi_n^{(0)} \rangle. \end{aligned} \tag{2.58}$$

The transition probability per unit of time is

$$w_{mn} = \frac{2\pi}{\hbar} |\langle \varphi_m | T(E_n) | \varphi_n \rangle|^2\, \delta(E_n - E_m)\,. \tag{2.59}$$

This expression is called “Fermi’s Golden Rule” [50,51].

The summation of all the final states results in

$$\begin{aligned} \sum_m w_{mn} &= \sum_m |\langle \varphi_m^{(0)} | T_1 | \varphi_n^{(0)} \rangle|^2 = \sum_m \langle \varphi_n^{(0)} | T_1^+ | \varphi_m^{(0)} \rangle \langle \varphi_m^{(0)} | T_1 | \varphi_n^{(0)} \rangle \\ &= \langle \varphi_n^{(0)} | T_1^+ T_1 | \varphi_n^{(0)} \rangle = -\langle \varphi_n^{(0)} | T_1^+ + T_1 | \varphi_n^{(0)} \rangle. \end{aligned} \tag{2.60}$$

The above derivation starts from Eq. (2.1). The Dirac equation is also of this form, so that the above formalism is valid for both the Schrödinger equation and Dirac equation. The formalism applies to three-, two-, and one-dimensions. In the following, the above generalized scattering theory is applied to one- and two-particle systems.

## 3. Singe-particle scattering

It is seen that the forms of the formulas in the generalized scattering theory are basically the same as those of the single-particle scattering theory introduced in the literature.

Now, we apply the generalized theory to the case of single-particle scattering. On one hand, this is the simplest example. On the other hand, we introduce the two scattering pictures usually presented in textbooks, clarifying the concept of the probability of the transition between the initial and final states.

The spatial coordinate set $R$ is that of a particle in three-dimensional space, $R \Rightarrow \boldsymbol{r}$. $H_0$ is the Hamiltonian of a free particle.

$$H_0 = -\frac{\hbar^2}{2m} \nabla^2\,. \tag{3.1}$$

The stationary wave function of the $H_0$ is

$$\varphi_{\boldsymbol{p}}^{(0)}(\boldsymbol{r})=\frac{\mathrm{e}^{\mathrm{i}\boldsymbol{p}\cdot\boldsymbol{r}/\hbar}}{(2\pi)^{3/2}}. \tag{3.2}$$

Corresponding eigenvalue is

$$E(\boldsymbol{p})=\frac{\boldsymbol{p}^2}{2m}. \tag{3.3}$$

The wave function obeying the free particle's Schrödinger equation is

$$\psi_{\boldsymbol{p}}^{(0)}(\boldsymbol{r},t)=\varphi_{\boldsymbol{p}}^{(0)}(\boldsymbol{r})\mathrm{e}^{-\mathrm{i}E(\boldsymbol{p})t/\hbar}=\frac{1}{(2\pi)^{3/2}}\mathrm{e}^{-\mathrm{i}(E(\boldsymbol{p})t-\boldsymbol{p}\cdot\boldsymbol{r})/\hbar}. \tag{3.4}$$

The Green's function of a single particle is denoted by a lowercase letter $g$. When Eqs. (3.3) and (3.4) are substituted into (2.23), we obtain a free article's Green's function [35]:

$$G^{(0)}=g^{(0)}(\boldsymbol{r},\boldsymbol{r}';t,t_0)=A\mathrm{e}^{\mathrm{i}(\boldsymbol{r}-\boldsymbol{r}')^2 m/2\hbar(t-t_0)}, \tag{3.5a}$$

where

$$A=\frac{m^{3/2}}{(2\pi\mathrm{i}\hbar(t-t_0))^{3/2}}. \tag{3.5b}$$

This Green's function is of the property (2.24). If the scattering potential $H_1$ meets (2.32), the $T$ follows (2.33).

The retarded Green's function is

$$g^{\mathrm{R}(0)}=\theta(t-t_0)g^{(0)}. \tag{3.6}$$

It is easily verified that the wave function (3.4) and Green's function (3.5) satisfy (2.19) and (2.26).

Suppose that there is a scattering center at the origin. A free particle is incident from far way. When it approaches the scattering center, it suffers a potential $H_1$. After scattering, the wave function can be evaluated by (2.37) which is in the present case as follows.

$$\psi^{\mathrm{R}}(\boldsymbol{r},t)=\psi^{\mathrm{R}(0)}(\boldsymbol{r},t)+\int_{-\infty}^{\infty}\mathrm{d}t_1\int\mathrm{d}\boldsymbol{r}_1 g^{\mathrm{R}(0)}(\boldsymbol{r},\boldsymbol{r}_1;t,t_1)T\psi^{\mathrm{R}(0)}(\boldsymbol{r}_1,t_1). \tag{3.7}$$

This is the time-dependent Lippmann-Schwinger equation [7]. It is just a special case of Eq. (2.37) when applied to single-particle scattering.

Now, assume that the incident wave is just (3.4), an eigenstate of the Hamiltonian (3.1).

$$\psi_n^{(0)}(t)=\psi_{\boldsymbol{p}}^{(0)}(\boldsymbol{r},t). \tag{3.8}$$

Equation (3.7) is more explicitly written as

$$\psi_{\boldsymbol{p}}(\boldsymbol{r},t)=\psi_{\boldsymbol{p}}^{(0)}(\boldsymbol{r},t)+\int_{t_0}^{t}\mathrm{d}t_1\int\mathrm{d}\boldsymbol{r}_1 g^{(0)}(\boldsymbol{r},\boldsymbol{r}_1;t,t_1)T\psi_{\boldsymbol{p}}^{(0)}(\boldsymbol{r}_1,t_1). \tag{3.9}$$

The $\psi_{\boldsymbol{p}}(\boldsymbol{r},t)$ on the left hand side is the final state evolved from the initial state $\psi_{\boldsymbol{p}}^{(0)}(\boldsymbol{r},t)$. Equation (3.9) is the special case of (2.43), and is the basis of evaluation of cross section.

In the following, we schematically show scattering by figures. In the figures, we use a straight line with an arrow and a bold letter to represent a wave function with a momentum. A solid line means the incident wave before scattering, and dashed lines mean outgoing waves after scattering.

We assume that the potential $H_1$ in (3.9) is independent of time. Equation (3.9) reflects the physical picture of scattering as in Fig. 1 where the wave functions of $H_0$ and $H_0 + H_1$ are illustrated. In Fig. 1, the arrowed solid line is the initial state, a plane wave with a momentum $\boldsymbol{p}$, which is the function (3.4). The dashed lines represent the final state, the eigen function of $H_0 + H_1$. The final state actually contains two parts, represented by thicker and thinner lines, respectively. The thicker dashed line is an outgoing plane wave, which is the first term in (3.9). The thinner lines are the second term in (3.9), which spread in all directions.

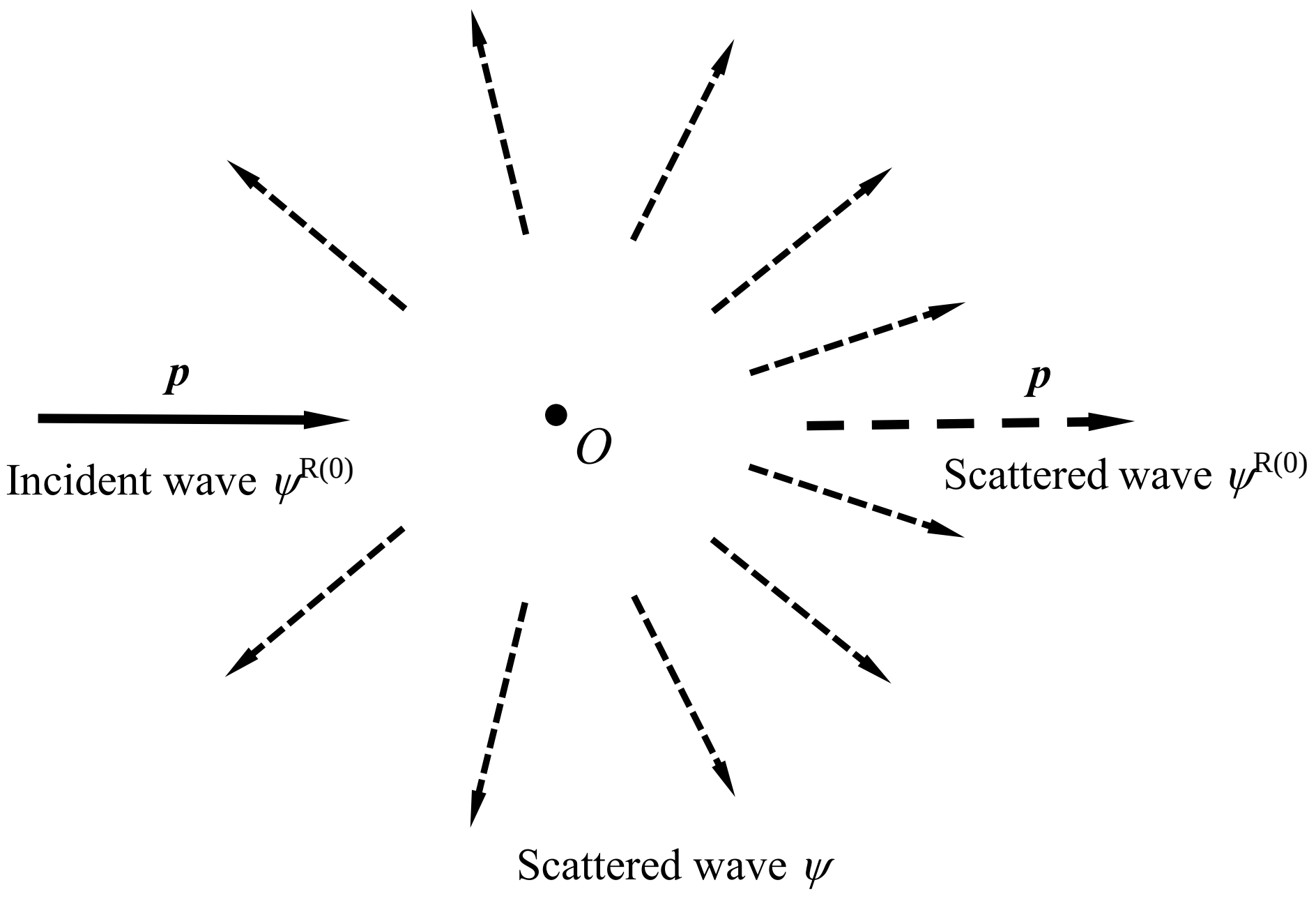

**Fig. 1.** The picture of single particle scattering [5,6,46,51–53]. The solid line means the incident wave $\psi^{R(0)}$ with momentum $\boldsymbol{p}$. The dashed lines mean the waves after scattering, which are composed of two parts: that exactly the same as the incident wave

$\psi^{R(0)}$ shown by thicker dashed line and scattered wave $\psi$ shown by thinner dashed lines, the second term in (3.9).

Besides Fig. 1, there is another picture describing the single-particle scattering, usually seen in textbooks of relativistic QM and quantum electrodynamics, shown in Fig. 2(a).

Figure 2(a) describes the scattering process in the following language. The initial state is a plane wave of an incident particle with a momentum $\boldsymbol{p}$. When the particle is far away from the origin, it does not feel interaction. When it is approaching the origin, the $H_1$ is turned on, and the particle suffers the interaction. After the scattering, the particle moves away from the origin. When it is far enough, it does not undergo the interaction anymore. That is to say, the $H_1$ is turned off, and the particle become a new plane wave with a momentum $\boldsymbol{p}'$, as shown by the arrowed dashed line in Fig. 2(a). According to this description, the potential $H_1$ varies with time. The initial and final states do not suffer interaction so that both are plane waves. Figure 2(a) is conceptually different from Fig. 1 in that the former means the Hamiltonian depends on time while the latter does not and in that the two final states are different.

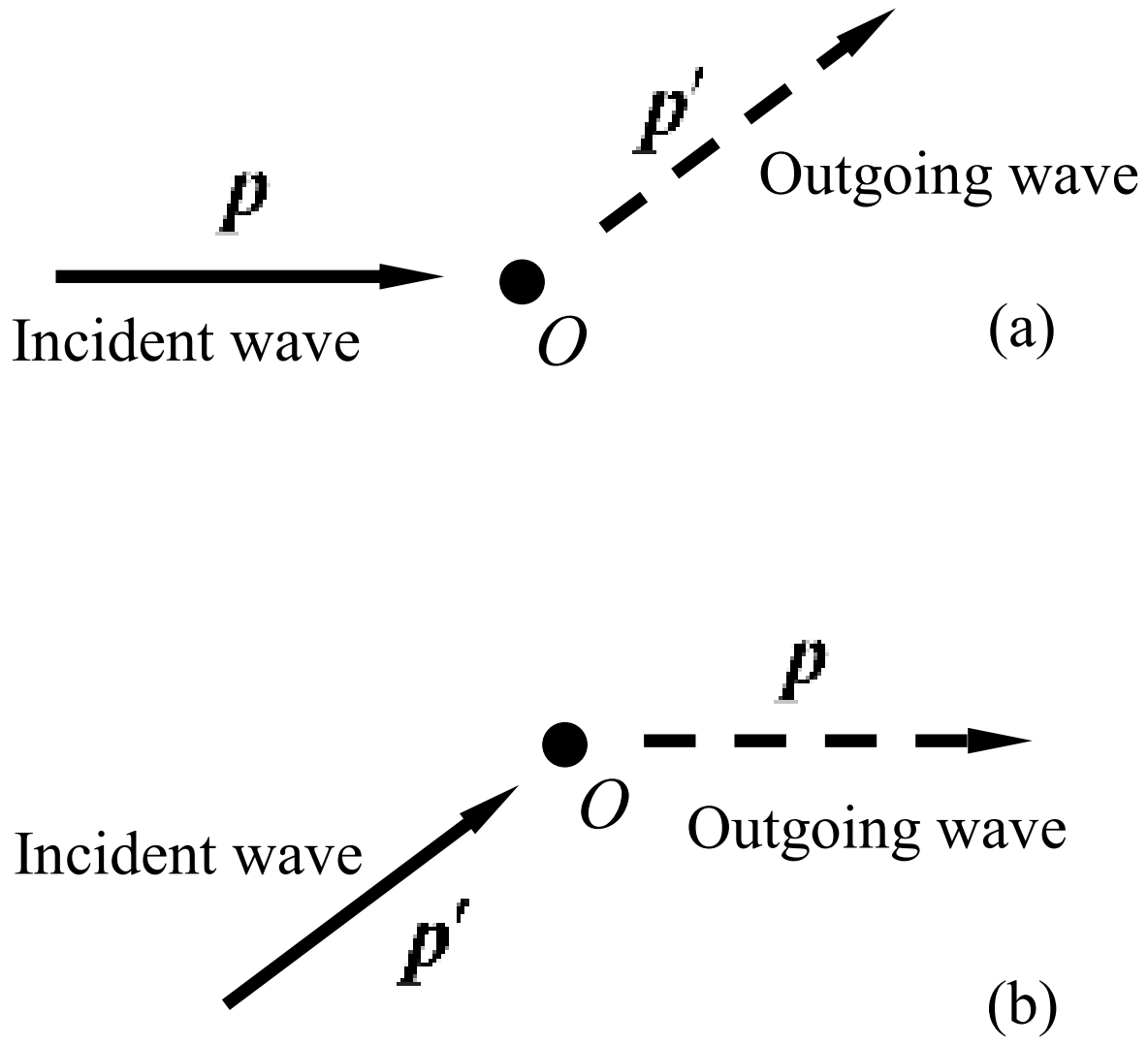

**Fig. 2.** An illustration of one-particle transition process and its reciprocal process. (a) A particle with momentum $\boldsymbol{p}$ suffers an interaction at the origin, and then, transits to a new plane wave with momentum $\boldsymbol{p}'$ [6]. Initial and final states, both being plane waves, are represented by the solid and dashed lines, respectively. (b) The inverse of

(a).

The author thinks that Fig. 1 describes the single-particle scattering correctly. Firstly, in Fig. 1, the dashed lines represent the wave function of (3.9), which is the solution of the Schrödinger equation with Hamiltonian *H*. The scattering wave spreads in all directions. This should be the correct physical picture of the scattering. In contrast, the plane wave $\boldsymbol{p}'$ in Fig. 2 does not satisfy the Schrödinger equation with Hamiltonian *H*. Secondly, the whole Hamiltonian of a gas in an equilibrium state is time-independent. The gas is composed of a large amount of molecules that collide with each other frequently. The Hamiltonian of a pair of particles colliding with each other should be independent of time because the whole Hamiltonian of the gas is.

According to the description of Fig. 2, the scattering Hamiltonian varies with time, so that it is a virtual Hamiltonian. The final state is a plane wave, which does not agree with (3.9).

Nevertheless, Fig. 2(a) is still useful in understanding transition probability. In fact, the plane waves (3.4) constitute a complete set. The scattering wave, the second term in (3.9), can be written as the linear superposition of all the plane waves. Figure 2(a) only shows one outgoing plane wave with a momentum $\boldsymbol{p}'$. Hence, what Fig. 2(a) shows is actually the projection of the final state onto a plane wave with momentum $\boldsymbol{p}'$. Besides this plane wave, the final state can also project onto other plane waves. On every plane wave, there is a projection probability.

The project amplitude is expressed by Eq. (2.44). We are considering elastic scattering. Substituting (3.8) and (3.9) into (2.44), we obtain

$$\begin{aligned} C(\boldsymbol{p},\boldsymbol{p}',t) &= \int \mathrm{d}\boldsymbol{r}\psi_{\boldsymbol{p}'}^{(0)*}(\boldsymbol{r},t)\psi_{\boldsymbol{p}}^{(0)}(\boldsymbol{r},t) \\ &= \delta(\boldsymbol{p}-\boldsymbol{p}') + \int_{t_0}^{t} \mathrm{d}t_1 \int \mathrm{d}\boldsymbol{r}'\varphi_{\boldsymbol{p}'}^{(0)*}(\boldsymbol{r}_1)T\varphi_{\boldsymbol{p}}^{(0)}(\boldsymbol{r}_1). \end{aligned} \tag{3.10}$$

This is the projection amplitude of the final state in Fig. 1 onto the outgoing plane wave $\boldsymbol{p}'$ in Fig. 2. It is also the transition amplitude of a particle from plane wave with $\boldsymbol{p}$ to that with $\boldsymbol{p}'$. Here, we have assumed that $H_1$ is time-independent.

If the *T* only takes the first-order term in (2.31), the first-order term in (3.10) is

$$C^{(1)}(\boldsymbol{p},\boldsymbol{p}') = \int \mathrm{d}\boldsymbol{r}\varphi_{\boldsymbol{p}'}^{(0)*}(\boldsymbol{r})H_1\varphi_{\boldsymbol{p}}^{(0)}(\boldsymbol{r}). \tag{3.11}$$

For a Coulomb potential,

$$H_1 = \frac{q_1 q}{4\pi\varepsilon_0 |\boldsymbol{r}|}, \tag{3.12}$$

where $q_1$ and *q* are the charges of the incident particle and scatter center, respectively.

Under this potential, the first-order transition amplitude will be

$$C^{(1)}(\boldsymbol{p},\boldsymbol{p}') = \frac{q_1 q}{4\pi\varepsilon_0(\boldsymbol{p}-\boldsymbol{p}')^2}. \tag{3.13}$$

Let us consider the inverse process of Fig. 2(a). When the $\boldsymbol{p}$ and $\boldsymbol{p}'$ are exchanged, the resultant is shown in Fig. 2(b). In Fig. 2(b), the plane wave $\boldsymbol{p}'$ is the incident wave, and the outgoing plane has a momentum $\boldsymbol{p}$. The Hamiltonians of Figs. 2(a) and (b) are the same. We want to calculate the transition amplitude of this inverse process. This is easily done because we merely need to exchange the $\boldsymbol{p}$ and $\boldsymbol{p}'$ in (3.10), and the result is

$$C(\boldsymbol{p},\boldsymbol{p}';t) = C^*(\boldsymbol{p}',\boldsymbol{p};t). \tag{3.14}$$

The transition amplitudes of this pair of reciprocal processes are complex conjugate to each other. Consequently, their transition probabilities are equal.

$$w(\boldsymbol{p},\boldsymbol{p}';t) = |C(\boldsymbol{p},\boldsymbol{p}',t)|^2 = w(\boldsymbol{p}',\boldsymbol{p};t). \tag{3.15a}$$

It is easily seen that when all the momenta above take opposite directions, Eqs. (3.14) and (3.15) still stand. That is to say,

$$w(-\boldsymbol{p},-\boldsymbol{p}';t) = w(-\boldsymbol{p}',-\boldsymbol{p};t). \tag{3.15b}$$

In the direction of the $\boldsymbol{p}'$, the differential cross section is proportional to the transition probability (3.15), $\frac{d\sigma}{d\Omega} \propto |C(\boldsymbol{p},\boldsymbol{p}',t)|^2$. For elastic scattering, it happens to be $\frac{d\sigma}{d\Omega} = |C(\boldsymbol{p},\boldsymbol{p}',t)|^2$.

There was a more concise formula to evaluation the differential cross section [6]:

$$\sigma(\theta,\varphi) = \frac{r^2 j_s(\theta,\varphi)}{j_0}, \tag{3.16}$$

where $j_0$ and $j_s$ are the currents of the incident and outgoing particles, respectively. In one- and two-dimensional spaces, Eq. (3.16) is respectively modified to be $\sigma = \frac{j_s}{j_0}$ and $\sigma(\theta) = \frac{r j_s(\theta)}{j_0}$. We [54] utilized these formulas to evaluate the differential cross sections up to the first order for both the low-momentum and relativistic particles in

three-, two-, and one-dimensional spaces.

## 4. Two-particle scattering

We take laboratory frame, in which both particles move before and after collision. In Fig. 3(a) illustrated is the collision process of two classical particles. We assume that the momenta $\boldsymbol{p}_1$ and $\boldsymbol{p}_2$ of the two particles before the scattering are known. After the scattering, their outgoing momenta are $\boldsymbol{p}_1'$ and $\boldsymbol{p}_2'$, respective. Please note that the outgoing momenta are not unique, which was discussed in [55]. Figure 3(b) illustrates the reciprocal or inverse process of Fig. 3(a).

We are investigating scattering problem in QM. The spatial coordinate set $R$ is the coordinates of two particles in three-dimensional space, $R \Rightarrow (\boldsymbol{r}_1, \boldsymbol{r}_2)$. $H_0$ is the Hamiltonian of two free particles.

$$H_0 = -\frac{\hbar^2}{2m_1}\frac{\partial^2}{\partial \boldsymbol{r}_1^2} - \frac{\hbar^2}{2m_2}\frac{\partial^2}{\partial \boldsymbol{r}_2^2}. \tag{4.1}$$

The total energy of the two free particles is

$$E(\boldsymbol{p}_1, \boldsymbol{p}_2) = \frac{\boldsymbol{p}_1^2}{2m_1} + \frac{\boldsymbol{p}_2^2}{2m_2}. \tag{4.2}$$

### 4.1 Distinguishable particles

The stationary equation of Hamiltonian (4.1) are as follows.

$$H_0 \varphi_{\boldsymbol{p}1}^{(0)}(\boldsymbol{r}_1)\varphi_{\boldsymbol{p}2}^{(0)}(\boldsymbol{r}_2) = E(\boldsymbol{p}_1, \boldsymbol{p}_2)\varphi_{\boldsymbol{p}1}^{(0)}(\boldsymbol{r}_1)\varphi_{\boldsymbol{p}2}^{(0)}(\boldsymbol{r}_2). \tag{4.3}$$

The two single-particle' stationary wave functions are

$$\varphi_{\boldsymbol{p}1}^{(0)}(\boldsymbol{r}_1) = \frac{1}{(2\pi)^{3/2}} \mathrm{e}^{\mathrm{i}\boldsymbol{p}_1\cdot\boldsymbol{r}_1/\hbar}, \varphi_{\boldsymbol{p}2}^{(0)}(\boldsymbol{r}_2) = \frac{1}{(2\pi)^{3/2}} \mathrm{e}^{\mathrm{i}\boldsymbol{p}_2\cdot\boldsymbol{r}_2/\hbar}. \tag{4.4}$$

A wave function satisfying Eq. (2.14a) is the product of two particles' wave functions (3.4).

$$\Psi_n^{(0)}(R,t) = \psi_{\boldsymbol{p}1}^{(0)}(\boldsymbol{r}_1,t)\psi_{\boldsymbol{p}2}^{(0)}(\boldsymbol{r}_2,t) = \varphi_{\boldsymbol{p}1}^{(0)}(\boldsymbol{r}_1)\varphi_{\boldsymbol{p}2}^{(0)}(\boldsymbol{r}_2)\mathrm{e}^{-\mathrm{i}E(\boldsymbol{p}_1,\boldsymbol{p}_2)t/\hbar} \tag{4.5}$$

The corresponding Green's function is

$$\begin{aligned} G^{(0)}(R,R';t,t_0) &= \sum_{\boldsymbol{p}1,\boldsymbol{p}2} \psi_{\boldsymbol{p}1}^{(0)}(\boldsymbol{r}_1,t)\psi_{\boldsymbol{p}2}^{(0)}(\boldsymbol{r}_2,t)\psi_{\boldsymbol{p}1}^{(0)*}(\boldsymbol{r}_1',t_0)\psi_{\boldsymbol{p}2}^{(0)*}(\boldsymbol{r}_2',t_0) \\ &= g^{(0)}(\boldsymbol{r}_1,\boldsymbol{r}_1';t,t_0)g^{(0)}(\boldsymbol{r}_2,\boldsymbol{r}_2';t,t_0), \end{aligned} \tag{4.6}$$

where the expression of the $g^{(0)}$ is Eq. (3.5). Equation (4.6) is simply the product of the two particles' Green's functions. It is of the property (2.24). The wave function (4.5) and Green's function (4.6) satisfy (2.19) and (2.26).

If the interaction $H_1$ between the two particles is real, then, the $T$ is also real. When the two particles are far away from each other, the interaction between them is weak enough so that they can be treated as free particles. When they are close to each other, the interaction $H_1$ takes effect. After the collision, the final state is evaluated by Eq. (2.43). Substituting (4.5) and (4.6) into (2.43), we obtain

$$\begin{aligned}\psi_n(\boldsymbol{r}_1,\boldsymbol{r}_2;t) &= \varphi^{(0)}_{\boldsymbol{p}1}(\boldsymbol{r}_1)\varphi^{(0)}_{\boldsymbol{p}2}(\boldsymbol{r}_2)\mathrm{e}^{-\mathrm{i}E(\boldsymbol{p}_1,\boldsymbol{p}_2)t/\hbar} \\ &+\int_{t_0}^{t}\mathrm{d}t_1\int\mathrm{d}\boldsymbol{r}_1'\int\mathrm{d}\boldsymbol{r}_2' g^{(0)}(\boldsymbol{r}_1,\boldsymbol{r}_1';t,t_1)g^{(0)}(\boldsymbol{r}_2,\boldsymbol{r}_2';t,t_1)T\varphi^{(0)}_{\boldsymbol{p}1}(\boldsymbol{r}_1')\varphi^{(0)}_{\boldsymbol{p}2}(\boldsymbol{r}_2')\mathrm{e}^{-\mathrm{i}E(\boldsymbol{p}_1,\boldsymbol{p}_2)t_1/\hbar}.\end{aligned} \tag{4.7}$$

In Fig. 3(a), two arrowed solid lines with letters $\boldsymbol{p}_1$ and $\boldsymbol{p}_2$ represent the plane waves before the collision. After the collision, the form of the final state $\psi^{\mathrm{R}}(\boldsymbol{r}_1,\boldsymbol{r}_2;t)$ is complex, because the scattering waves of the two particles spread in all directions, as may be imagined referring to Fig. 1. We merely consider the amplitude that the two particles transit, after the collision, to two plane waves with momenta $\boldsymbol{p}_1'$ and $\boldsymbol{p}_2'$, respectively. The transition is sketched in Fig. 3(a). That is to say, what we want to compute is the projection of the final state $\psi_n^{\mathrm{R}}(\boldsymbol{r}_1,\boldsymbol{r}_2;t)$ onto wave function

$$\Psi^{(0)}_m(R,t) = \psi^{(0)}_{\boldsymbol{p}'1}(\boldsymbol{r}_1,t)\psi^{(0)}_{\boldsymbol{p}'2}(\boldsymbol{r}_2,t). \tag{4.8}$$

Using Eqs. (4.7) and (4.8) in (2.44), the transition amplitude is

$$\begin{aligned}C(\boldsymbol{p}_1,\boldsymbol{p}_2;\boldsymbol{p}_1',\boldsymbol{p}_2';t) &= \delta(\boldsymbol{p}_1-\boldsymbol{p}_1')\delta(\boldsymbol{p}_2-\boldsymbol{p}_2') \\ &+\int_{t_0}^{t}\mathrm{d}t_1\int\mathrm{d}\boldsymbol{r}_1\int\mathrm{d}\boldsymbol{r}_2\varphi^{(0)}_{\boldsymbol{p}'1}(\boldsymbol{r}_1)\varphi^{(0)}_{\boldsymbol{p}'2}(\boldsymbol{r}_2)T\varphi^{(0)}_{\boldsymbol{p}1}(\boldsymbol{r}_1)\varphi^{(0)}_{\boldsymbol{p}2}(\boldsymbol{r}_2).\end{aligned} \tag{4.9}$$

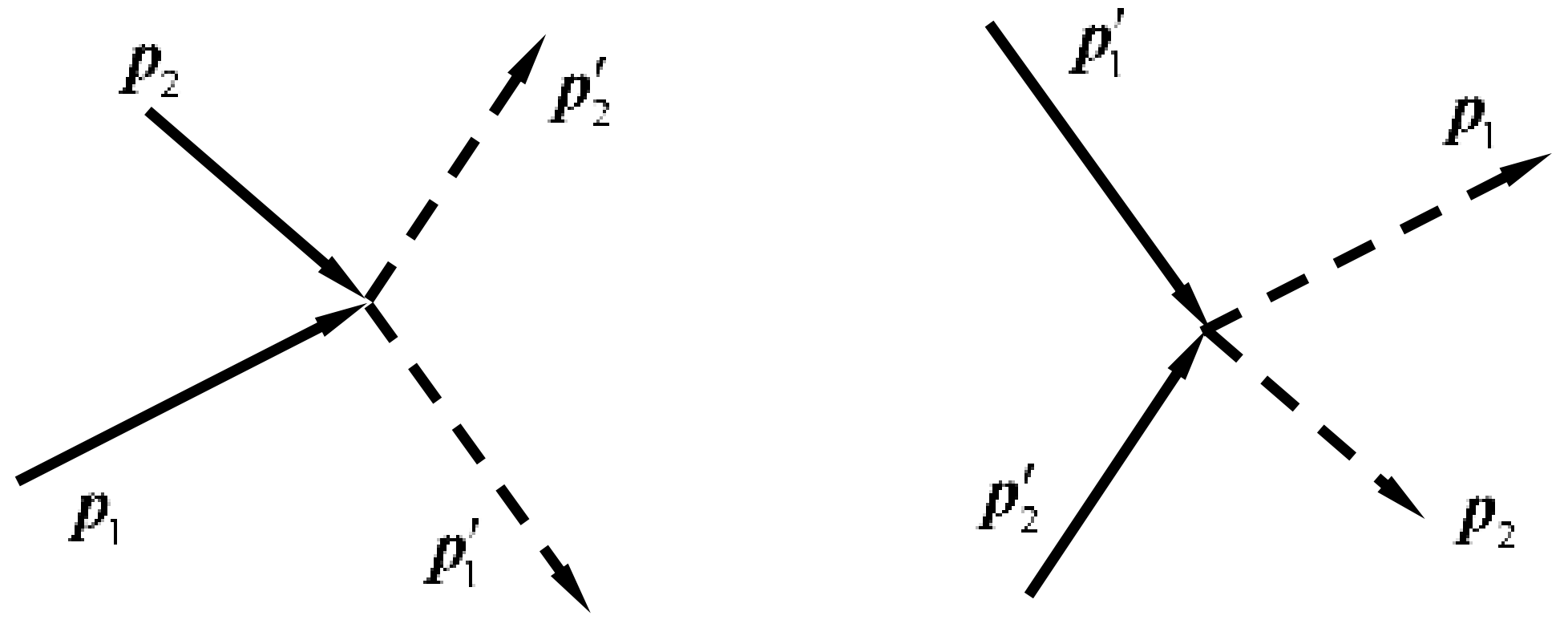

**(a)** **(b)**

**Fig. 3.** (a) Illustration of two-particle collision. In classical mechanics, two particles with momenta $\boldsymbol{p}_1$ and $\boldsymbol{p}_2$ collide, and after that, their momenta become $\boldsymbol{p}_1'$ and $\boldsymbol{p}_2'$, respectively. Only one pair of possible outgoing momenta is depicted. In QM, two plane waves with momenta $\boldsymbol{p}_1$ and $\boldsymbol{p}_2$ collide and then transit to two new plane waves with momenta $\boldsymbol{p}_1'$ and $\boldsymbol{p}_2'$. Any pair of $\boldsymbol{p}_1'$ and $\boldsymbol{p}_2'$ that satisfies the momentum and kinetic energy conservations is possible. (b) The inverse process of (a) [43,44,56–59].

We consider the inverse process of Fig. 3(a) [43,44,56–59]. That is to say, the $\boldsymbol{p}_1$ and $\boldsymbol{p}_1'$ are exchanged and $\boldsymbol{p}_2$ and $\boldsymbol{p}_2'$ exchanged, as depicted in Fig. 3(b). Before collision, the two particles have respectively momenta $\boldsymbol{p}_1'$ and $\boldsymbol{p}_2'$, and after collision, their momenta become $\boldsymbol{p}_1$ and $\boldsymbol{p}_2$. The wave function of the final state is evaluated by simply replacing the $\boldsymbol{p}_1$ and $\boldsymbol{p}_2$ in (4.7) by $\boldsymbol{p}_1'$ and $\boldsymbol{p}_2'$. The projection amplitude of the final state onto the state (4.5) is just the amplitude that the plane waves $\boldsymbol{p}_1'$ and $\boldsymbol{p}_2'$ transit to $\boldsymbol{p}_1$ and $\boldsymbol{p}_2$. This transition amplitude $C(\boldsymbol{p}_1', \boldsymbol{p}_2'; \boldsymbol{p}_1, \boldsymbol{p}_2; t)$ is easily calculated simply by exchanging $\boldsymbol{p}_1$ and $\boldsymbol{p}_1'$ and exchanging $\boldsymbol{p}_2$ and $\boldsymbol{p}_2'$ in (4.9). Then, we obtain

$$C(\boldsymbol{p}_1', \boldsymbol{p}_2'; \boldsymbol{p}_1, \boldsymbol{p}_2; t) = C^*(\boldsymbol{p}_1, \boldsymbol{p}_2; \boldsymbol{p}_1', \boldsymbol{p}_2'; t). \tag{4.10}$$

Subsequently,

$$w(\boldsymbol{p}_1, \boldsymbol{p}_2, \boldsymbol{p}_1', \boldsymbol{p}_2'; t) = |C(\boldsymbol{p}_1, \boldsymbol{p}_2, \boldsymbol{p}_1', \boldsymbol{p}_2'; t)|^2 = w(\boldsymbol{p}_1', \boldsymbol{p}_2', \boldsymbol{p}_1, \boldsymbol{p}_2; t). \tag{4.11a}$$

This is just Eq. (1.8). The pair of reciprocal processes sketched by Figs. 3(a) and (b) have the same transition probability. Furthermore, if all the momenta in (4.11a) take opposite directions, we acquire

$$w(-\boldsymbol{p}_1, -\boldsymbol{p}_2, -\boldsymbol{p}_1', -\boldsymbol{p}_2'; t) = w(-\boldsymbol{p}_1', -\boldsymbol{p}_2', -\boldsymbol{p}_1, -\boldsymbol{p}_2; t). \tag{4.11b}$$

In Section Introduction, we have mentioned that the quantity $w(\boldsymbol{p}_1, \boldsymbol{p}_2, \boldsymbol{p}_1', \boldsymbol{p}_2'; t)$ is the one that is used in establishing Boltzmann transport equation in statistical mechanics. In the literature, Eq. (1.8) is a postulation. Here, we prove it. The key is that we give the formula evaluating the transition amplitude of two-particle scattering in laboratory frame. Equation (1.8) is hardly proved if the problem of two-particle scattering is reduced to one-particle scattering in mass-center frame.

As a matter of fact, in Section 2, we have represented a universal formula showing that any scattering transition and its reciprocal process have the same transition probability for any number of particle. Suppose that $N$ free particles collide. After the collision, the final state is projected onto a new set of $N$ plane waves with momenta different from those before the collision. The formula of the projection amplitude was given, and then, the equality of a pair of reciprocal processes of scattering was proven, as manifested by Eqs. (2.44)-(2.50).

When the interaction between two particles is

$$H_1 = \frac{q_1 q_2}{4\pi\varepsilon_0 |\boldsymbol{r}_1 - \boldsymbol{r}_2|}, \tag{4.12}$$

the first-order transition amplitude, by (4.9), will be

$$\begin{aligned} C^{(1)}(\boldsymbol{p}_1, \boldsymbol{p}_2; \boldsymbol{p}_1', \boldsymbol{p}_2') &= \int \mathrm{d}\boldsymbol{r}_1' \int \mathrm{d}\boldsymbol{r}_2' \varphi_{\boldsymbol{p}'1}^{(0)}(\boldsymbol{r}_1') \varphi_{\boldsymbol{p}'2}^{(0)}(\boldsymbol{r}_2') H_1 \varphi_{\boldsymbol{p}1}^{(0)}(\boldsymbol{r}_1') \varphi_{\boldsymbol{p}2}^{(0)}(\boldsymbol{r}_2') \\ &= \frac{q_1 q_2}{4\pi\varepsilon_0 (\boldsymbol{p}_1 - \boldsymbol{p}_1')^2} \delta(\boldsymbol{p}_1 - \boldsymbol{p}_1' + \boldsymbol{p}_2 - \boldsymbol{p}_2'). \end{aligned} \tag{4.13}$$

In the following discussion, we omit the time integrals in (4.9).

Let us consider a detector in the $(\varphi, \theta)$ direction. The particle current it detects is proportional to the transition probability by which the outgoing particles move in this direction.

Since the scattering is elastic, the total momentum and kinetic energy of the two particles are conserved.

$$\boldsymbol{p}_1 + \boldsymbol{p}_2 = \boldsymbol{p}_1' + \boldsymbol{p}_2'. \tag{4.14a}$$

$$\frac{1}{2m_1} \boldsymbol{p}_1^2 + \frac{1}{2m_2} \boldsymbol{p}_2^2 = \frac{1}{2m_1} \boldsymbol{p}_1'^2 + \frac{1}{2m_2} \boldsymbol{p}_2'^2. \tag{4.14b}$$

If $\boldsymbol{p}_1'$ is along the $(\varphi, \theta)$ direction, the unknown quantities will be the absolute value of $\boldsymbol{p}_1'$, $|\boldsymbol{p}_1'| = p_1'$ and the three coordinates of the $\boldsymbol{p}_2'$. The four unknown quantities can be solved from Eq. (4.14). We denote $\boldsymbol{p}_1'$ and $\boldsymbol{p}_2'$ as $\boldsymbol{p}_1'(\varphi, \theta)$ and $\boldsymbol{p}_2'(\boldsymbol{p}_1'(\varphi, \theta))$, respectively. The amplitude that the two particles transit to this state is

$$\begin{aligned} &C_a(\boldsymbol{p}_1, \boldsymbol{p}_2; \theta, \varphi) \\ &= \int \mathrm{d}\boldsymbol{r}_1 \mathrm{d}\boldsymbol{r}_2 \varphi^{(0)}(\boldsymbol{p}_1'(\theta, \varphi), \boldsymbol{r}_1) \varphi^{(0)}(\boldsymbol{p}_2'(\boldsymbol{p}_1'(\theta, \varphi)), \boldsymbol{r}_2) T \varphi_{\boldsymbol{p}_1}^{(0)}(\boldsymbol{r}_1) \varphi_{\boldsymbol{p}_2}^{(0)}(\boldsymbol{r}_2). \end{aligned} \tag{4.15a}$$

On the other hand, $\boldsymbol{p}_2'$ can also be in the $(\varphi, \theta)$ direction. Correspondingly, the four unknown quantities are $|\boldsymbol{p}_2'| = p_2'$ and the three coordinates of the $\boldsymbol{p}_1'$, which can be solved from (4.14). We denote the $\boldsymbol{p}_1'$ and $\boldsymbol{p}_2'$ in this case as $\boldsymbol{p}_2'(\varphi,\theta)$ and $\boldsymbol{p}_1'(\boldsymbol{p}_2'(\varphi,\theta))$, respectively. The amplitude that the two particles transit to this state is

$$C_b(\boldsymbol{p}_1,\boldsymbol{p}_2;\theta,\varphi)$$
$$=\int \mathrm{d}\boldsymbol{r}_1\mathrm{d}\boldsymbol{r}_2\varphi^{(0)}(\boldsymbol{p}_1'(\boldsymbol{p}_2'(\theta,\varphi)),\boldsymbol{r}_1)\varphi^{(0)}(\boldsymbol{p}_2'(\theta,\varphi),\boldsymbol{r}_2)T\varphi_{\boldsymbol{p}_1}^{(0)}(\boldsymbol{r}_1)\varphi_{\boldsymbol{p}_2}^{(0)}(\boldsymbol{r}_2). \tag{4.15b}$$

The total amplitude is

$$C(\boldsymbol{p}_1,\boldsymbol{p}_2;\theta,\varphi)=C_a(\boldsymbol{p}_1,\boldsymbol{p}_2;\theta,\varphi)+C_b(\boldsymbol{p}_1,\boldsymbol{p}_2;\theta,\varphi). \tag{4.15c}$$

The particle current density detected by a detector in the $(\varphi,\theta)$ direction is proportional to

$$|C(\boldsymbol{p}_1,\boldsymbol{p}_2;\theta,\varphi;t)|^2. \tag{4.16}$$

Please note that there is a term of interference between the $C_a$ and $C_b$. As for the definition of differential cross section in this case, an example can be seen in [12].

### 4.2 Indistinguishable particles

For indistinguishable particles, the wave function (4.5) is not valid, and exchange effect should be taken into account. Oppenheimer first noticed that in treating scattering problems, the effect of identity of particles should be concerned [60]. The wave function of two identical particles is

$$\Psi_n^{(0)}(\boldsymbol{r}_1,\boldsymbol{r}_2,t)=\frac{1}{\sqrt{2}}[\varphi_{\boldsymbol{p}1}^{(0)}(\boldsymbol{r}_1)\varphi_{\boldsymbol{p}2}^{(0)}(\boldsymbol{r}_2)\pm\varphi_{\boldsymbol{p}1}^{(0)}(\boldsymbol{r}_2)\varphi_{\boldsymbol{p}2}^{(0)}(\boldsymbol{r}_1)]\mathrm{e}^{-\mathrm{i}E(\boldsymbol{p}_1,\boldsymbol{p}_2)t/\hbar}, \tag{4.17}$$

where the upper (lower) sign corresponds to spatial symmetric (antisymmetric) wave function. Here, spin is not taken into account. Substitution of (4.17) into (2.23) leads to

$$G^{(0)}=g^{(0)}(\boldsymbol{r}_1,\boldsymbol{r}_1')g^{(0)}(\boldsymbol{r}_2,\boldsymbol{r}_2')\pm g^{(0)}(\boldsymbol{r}_1,\boldsymbol{r}_2')g^{(0)}(\boldsymbol{r}_2,\boldsymbol{r}_1'), \tag{4.18}$$

where $g^{(0)}$ is (3.5). Compared to (4.6), Eq. (4.18) has one more term which embodies exchange effect. Equations (4.17) and (4.18) satisfy both (2.19) and (2.26). When (4.17) and (4.18) are substituted into (2.43), the final state can be written.

We evaluate the transition amplitude of the process shown in Fig. 3(a). This is the projection of the final state onto the following wave function:

$$\Psi_m^{(0)}(\boldsymbol{r}_1,\boldsymbol{r}_2,t)=\frac{1}{\sqrt{2}}[\varphi_{\boldsymbol{p}'1}^{(0)}(\boldsymbol{r}_1)\varphi_{\boldsymbol{p}'2}^{(0)}(\boldsymbol{r}_2)\pm\varphi_{\boldsymbol{p}'1}^{(0)}(\boldsymbol{r}_2)\varphi_{\boldsymbol{p}'2}^{(0)}(\boldsymbol{r}_1)]\mathrm{e}^{-\mathrm{i}E(\boldsymbol{p}_1',\boldsymbol{p}_2')t/\hbar}. \tag{4.19}$$

The projection amplitude is

$$\begin{aligned}
&C(\boldsymbol{p}_1,\boldsymbol{p}_2,\boldsymbol{p}_1',\boldsymbol{p}_2';t)=\int \mathrm{d}\boldsymbol{r}_1\int \mathrm{d}\boldsymbol{r}_2\Psi_m^{(0)*}(\boldsymbol{r}_1,\boldsymbol{r}_2,t)\psi_n^{\mathrm{R}}(\boldsymbol{r}_1,\boldsymbol{r}_2;t)\\
&=\delta(\boldsymbol{p}_1-\boldsymbol{p}_1')\delta(\boldsymbol{p}_2-\boldsymbol{p}_2')\pm\delta(\boldsymbol{p}_1-\boldsymbol{p}_2')\delta(\boldsymbol{p}_2-\boldsymbol{p}_1')\\
&+\int_{t_0}^{t}\mathrm{d}t_1\int \mathrm{d}\boldsymbol{r}_1\int \mathrm{d}\boldsymbol{r}_2\Psi_m^{(0)*}(\boldsymbol{r}_1,\boldsymbol{r}_2,t')T\Psi_n^{(0)}(\boldsymbol{r}_1,\boldsymbol{r}_2,t').
\end{aligned}\tag{4.20}$$

For the reciprocal process Fig. 3(b), the transition amplitude $C(\boldsymbol{p}_1,\boldsymbol{p}_2,\boldsymbol{p}_1',\boldsymbol{p}_2';t)$ can be easily written, as long as in (4.20) the $(\boldsymbol{p}_1,\boldsymbol{p}_2)$ and $(\boldsymbol{p}_1',\boldsymbol{p}_2')$ are exchanged. Consequently, Eq. (4.10) remains valid, and (4.11) as well.

The particle density detected by a detector in the $(\varphi,\theta)$ direction is proportional to the probability of transition to this direction. If $\boldsymbol{p}_1'$ is in the $(\varphi,\theta)$ direction, the $\boldsymbol{p}_1'$ and $\boldsymbol{p}_2'$ are denoted by $\boldsymbol{p}_1'(\varphi,\theta)$ and $\boldsymbol{p}_2'(\boldsymbol{p}_1'(\varphi,\theta))$, respectively. Then, the transition amplitude of the final state onto the wave function

$$\begin{aligned}
&\Psi_m^{(0)}(\boldsymbol{p}_1'(\theta,\varphi),\boldsymbol{p}_2'(\boldsymbol{p}_1'(\theta,\varphi));\boldsymbol{r}_1,\boldsymbol{r}_2,t)\\
&=\frac{1}{\sqrt{2}}[\varphi^{(0)}(\boldsymbol{p}_1'(\theta,\varphi),\boldsymbol{r}_1)\varphi^{(0)}(\boldsymbol{p}_2'(\boldsymbol{p}_1'(\theta,\varphi)),\boldsymbol{r}_2)\\
&\pm\varphi^{(0)}(\boldsymbol{p}_1'(\theta,\varphi),\boldsymbol{r}_2)\varphi^{(0)}(\boldsymbol{p}_2'(\boldsymbol{p}_1'(\theta,\varphi)),\boldsymbol{r}_1)]\mathrm{e}^{-\mathrm{i}E(\boldsymbol{p}_1',\boldsymbol{p}_2')t/\hbar}
\end{aligned}\tag{4.21a}$$

is denoted by $C_a(\boldsymbol{p}_1'(\theta,\varphi),\boldsymbol{p}_2'(\boldsymbol{p}_1'(\theta,\varphi)))$. On the other hand, if $\boldsymbol{p}_2'$ is in the $(\varphi,\theta)$ direction, the $\boldsymbol{p}_2'$ and $\boldsymbol{p}_1'$ in this case are denoted by $\boldsymbol{p}_2'(\varphi,\theta)$ and $\boldsymbol{p}_1'(\boldsymbol{p}_2'(\varphi,\theta))$, respectively. Then, the transition amplitude of the final state onto the wave function

$$\begin{aligned}
&\Psi_m^{(0)}(\boldsymbol{p}_2'(\theta,\varphi),\boldsymbol{p}_1'(\boldsymbol{p}_2'(\theta,\varphi));\boldsymbol{r}_1,\boldsymbol{r}_2,t)\\
&=\frac{1}{\sqrt{2}}[\varphi^{(0)}(\boldsymbol{p}_1'(\boldsymbol{p}_2'(\theta,\varphi)),\boldsymbol{r}_1)\varphi^{(0)}(\boldsymbol{p}_2'(\theta,\varphi),\boldsymbol{r}_2)\\
&\pm\varphi^{(0)}(\boldsymbol{p}_2'(\theta,\varphi),\boldsymbol{r}_1)\varphi^{(0)}(\boldsymbol{p}_1'(\boldsymbol{p}_2'(\theta,\varphi)),\boldsymbol{r}_2)]\mathrm{e}^{-\mathrm{i}E(\boldsymbol{p}_1',\boldsymbol{p}_2')t/\hbar}
\end{aligned}\tag{4.21b}$$

is denoted by $C_b(\boldsymbol{p}_1'(\boldsymbol{p}_1'(\theta,\varphi)),\boldsymbol{p}_2'(\theta,\varphi))$. The particle density detected in the $(\varphi,\theta)$ direction is proportional to $|C_a(\boldsymbol{p}_1'(\theta,\varphi),\boldsymbol{p}_2'(\boldsymbol{p}_1'(\theta,\varphi)))+C_b(\boldsymbol{p}_1'(\boldsymbol{p}_2'(\theta,\varphi)),\boldsymbol{p}_2'(\theta,\varphi))|^2$. Please note that we should not say that (4.21b) is the exchange term of (4.21a). This is because both $C_a(\boldsymbol{p}_1'(\theta,\varphi),\boldsymbol{p}_2'(\boldsymbol{p}_1'(\theta,\varphi)))$ and $C_b(\boldsymbol{p}_1'(\boldsymbol{p}_2'(\theta,\varphi)),\boldsymbol{p}_2'(\theta,\varphi))$ are specific cases of (4.20) which itself has contained exchange effect already.

The application of three-particle collision will be investigated later. Here, we merely mention that the triad of three-particle Lippmann-Schwinger equations [25-27,29] were actually from flexible divisions of $H_0$ and $H_1$ in the total three-particle Hamiltonian $H$. Each division lead to an equation, and the equations were combined to be solved.

## 5. Conclusion

In this work, a generalized scattering theory in quantum mechanics is established. The generalized Lippmann-Schwinger equation and scattering matrix are presented. Although the generalized formulas seem formally the same as those of original Lippmann-Schwinger theory, the physical implications are not the same. The generalized formalism can be applied to scattering problems of arbitrary number of particle. We stress that the derivation is rigorous, even for infinitesimals, without manual operation such as analytical continuation.

When this generalized theory is applied to single-particle scattering problem, the results usually introduced in quantum textbooks are retrieved, e.g., Lippmann-Schwinger equation and $S$ matrix. We clarified the concepts of the final state after scattering and transition.

When this generalized theory is applied to two-particle scattering, the formula of two-particle transition probability is derived. This probability emerged in the collision term in Boltzmann transport equation. It is proved that a pair of reciprocal scattering processes have the same transition probability. For the collision between two identical particles, the formula of transition probability is also presented.

Suppose that there are $N$ free particles described by plane waves. They collide and transit to a new set of plane waves. The transition probability of this scattering process is the same as that of the reciprocal process.

This work provides a useful means for dealing with the problems of many-particle scattering. Its realistic meaning will be shown in the author's following works. Here, we give two clues.

One clue is related to quantum electrodynamics (QED) where various scattering problems of particles are investigated. Usually, the famous Feynman propagator is used in evaluating scattering cross section. The Feynman propagator was obtained by analytic continuation, while the retarded Green's function in the present work is solved from a differential equation. There is a bit of difference between them (please compare Eqs. (43) and (53) in [35]). We will enter QED. In a previous work [54], the Green's functions were given, and the cross sections were evaluated up to the first-order term. In later works, we will revisit the topics in QED by means of the generalized scattering formulas presented in this paper, and clarify the role of the negative kinetic energy (NKE) solutions of relativistic quantum mechanics equations. Dirac explained the NKE solutions as the holes in filled electron sea. We have pointed out [61] that this explanation lead to contradictions. Feynman explained the NKE solutions as antiparticles moving counterclockwise. We will elaborate the physical picture of the NKE solutions in scattering processes. In the appendix D in [62], we listed 13 points that were topics to be studied. All of them, except point 12, have been dealt with in the author's works [61,63-71]. Point 12 will be touched later.

The other clue is related to the irreversibility of motion. It has been a long time that people felt puzzled by a paradox: every motion happened in reality is irreversible but the equations that reflect physical laws, say, Eq. (2.1), are time-reversible. We will figure out this problem. We will tell why all the happened motion is time-irreversible. Here, we merely mention the discrepancy of Eqs. (2.1) and (2.2): the former is of invariance of time-inversion but the latter is not. In the present work, we have shown that the transition probabilities of a pair of reciprocal scattering processes are the same. In terms of this result, we will tell the real meaning of the “detailed balance”.

**Acknowledgments**

This work is supported by the National Natural Science Foundation of China under Grant No. 12234013, and the National Key Research and Development Program of China Nos. 2018YFB0704304 and 2016YFB0700102.

**Appendix A. The meaning of the “manual operation”**

In Ref. [7], the author started from the Schrodinger equation, (1.1) in [7], which is copied here.

$$\mathrm{i}\hbar\frac{\partial}{\partial t}\Psi(t)=(H_0+H_1)\Psi(t) \tag{A1}$$

To solve this equation, there should be an initial condition, which sets the value of the wave function $\Psi'(t)$ at initial time $t_0$,

$$\Psi(t_0)=\Psi(t=t_0). \tag{A2}$$

However, they did not use an initial condition. Because of this, they actually were unable to directly solve (A1). They instead imposed a restriction (1.29):

$$V(\infty)+V(-\infty)=2. \tag{A3}$$

The function $V(t)$ was defined by

$$V(t)=\frac{2U_+(t)}{1+S}=\frac{2U_-(t)}{1+S^{-1}}, \tag{A4}$$

where the $U_+(t)$ and $U_-(t)$ were respectively time evolution operator and inverse time evolution operator, and *S* meant the *S* matrix. With the help of (A3), they obtained the solutions of (A1):

$$\Psi_a^{(\pm)}=\Phi_a+\frac{1}{E_a\pm\mathrm{i}\varepsilon-H_0}H_1\Psi_a^{(\pm)}, \tag{A5}$$

where $E_a$ and $\Phi_a$ are respectively the eigenenergy and eigen wave function of the $H_0$, i.e., $H_0\Phi_a = E_a\Phi_a$. Equation (A5) is the famous Lippmann-Schwinger equation.

Equation (A3) was called a boundary condition [7]. This so-called boundary condition is questionable because of the following reasons. The first reason is that Eq. (A1) is expected to be solved under an initial condition (A2) but not any other condition. No one has proved that in solving the differential equation (A1), the initial condition could be replaced by (A3). The meaning of the so-called boundary condition is unclear. The second one is that an initial condition can be set at any time $t_0$, but in (A3) the time is fixed at extreme points. The third one is that if one solves a time-retarded (time-advanced) solution, the initial condition should be irrelative to the information at time $t_0 \to +\infty$ ($t_0 \to -\infty$). However, the condition (A3) combines the information of both times $+\infty$ and $-\infty$. That is strange. Moreover, the *S* matrix should be evaluated by the solved wave function, but it had entered the condition (A3) before the wave function was known.

The derivation process of the Lippmann-Schwinger equation in [7] was too cumbersome, so it caused some puzzle. Some ones thought that the solution obtained might not be unique [18,19]. It was shown that the Lippmann-Schwinger equation was lacking from a rigorous mathematical basis, and its validity was still to be settled [27]. "The proper interpretation of the implied limit ε→0 in the equation is not obvious; indeed there are at least three plausible alternatives." [29]

Because the derivation process in [7] was an oblique routine, others choose a concise way. In [8,11], the authors directly insert a factor $\mathrm{e}^{\varepsilon t}$ into the integral expression of the solution, where $\varepsilon^{-1}$ was allowed to approach +∞, and was misunderstood as an energy shift. A simplest way, which is equivalent to the insertion of the factor $\mathrm{e}^{\varepsilon t}$, is to make analytic continuation, say, $E_a \to E_a \pm \mathrm{i}\varepsilon$ so as to obtain (A5) [14,17,18,19,21,30].

Now, we show that after the analytic continuation, the solution will be changed to be a form that does not satisfy the original differential equation.

Suppose that we want to solve the following differential equation.

$$(\mathrm{i}\hbar\frac{\partial}{\partial t}+\frac{\hbar^2}{2m}\nabla_r^2)G(\boldsymbol{r},\boldsymbol{r}_0;t,t_0)=\mathrm{i}\hbar\delta(\boldsymbol{r}-\boldsymbol{r}_0)\delta(t-t_0). \quad \text{(A6)}$$

Directly solving this equation gets the solution Eq. (3.5) which, in turn, can be verified to satisfy (A6).

Making use of Fourier transformations is another way to solve Eq. (A6). Let us implement time and space Fourier transformations.

$$G(\boldsymbol{r},\boldsymbol{r}_0;t,t_0)=\frac{1}{2\pi}\int \mathrm{d}\omega \mathrm{e}^{-\mathrm{i}\omega(t-t_0)}g(\boldsymbol{r},\boldsymbol{r}_0;\omega). \quad \text{(A7)}$$

$$g(\boldsymbol{r},\boldsymbol{r}_0;\omega)=\int\frac{\mathrm{d}\boldsymbol{k}}{(2\pi)^3}\mathrm{e}^{\mathrm{i}\boldsymbol{k}\cdot(\boldsymbol{r}-\boldsymbol{r}_0)}g(\boldsymbol{k};\omega). \tag{A8}$$

Substituting (A7) and (A8) into (A6), we achieve

$$g(\boldsymbol{k};\omega)=\frac{\mathrm{i}\hbar}{\hbar\omega-\hbar^2\boldsymbol{k}^2/2m}. \tag{A9}$$

In order to verify that (A9) satisfies (A6), we have to make inverse Fourier transformations. However, we find this cannot be done, because the poles in (A9) are in the real axis, which prevents us employing the residue theorem.

In order to conduct the inverse Fourier transformations, one takes analytic continuation $\omega\to\omega+\mathrm{i}0^+$:

$$g(\boldsymbol{k};\omega)\to g^+(\boldsymbol{k};\omega)=g(\boldsymbol{k};\omega+\mathrm{i}0^+)=\frac{\mathrm{i}\hbar}{\hbar\omega+\mathrm{i}0^+-\hbar^2\boldsymbol{k}^2/2m}. \tag{A10}$$

The inverse time Fourier transformation is

$$g^{\mathrm{R}}(\boldsymbol{k};t-t_0)=\int\frac{\mathrm{d}\omega}{2\pi}\mathrm{e}^{-\mathrm{i}\omega(t-t_0)}g^+(\boldsymbol{k};\omega)=\theta(t-t_0)\mathrm{e}^{-\mathrm{i}\hbar\boldsymbol{k}^2(t-t_0)/2m}. \tag{A11}$$

Subsequently, the inverse space Fourier transformation is

$$\begin{aligned}G^{\mathrm{R}}(\boldsymbol{r}-\boldsymbol{r}_0;t-t_0)&=\int\frac{\mathrm{d}\boldsymbol{k}}{(2\pi)^3}\mathrm{e}^{\mathrm{i}\boldsymbol{k}\cdot(\boldsymbol{r}-\boldsymbol{r}_0)}g^{\mathrm{R}}(\boldsymbol{k};t-t_0)\\&=\theta(t-t_0)\Big(\frac{m}{4\mathrm{i}\pi\hbar(t-t_0)}\Big)^{3/2}\exp\Big[\mathrm{i}\frac{m(\boldsymbol{r}-\boldsymbol{r}_0)^2}{2\hbar(t-t_0)}\Big]\end{aligned}. \tag{A12}$$

Equation (A12) does not meet (A6). As a matter of fact, it meets the following differential equation, as revealed in [35],

$$\Big(\mathrm{i}\hbar\frac{\partial}{\partial t}+\frac{\hbar^2}{2m}\nabla_r^2\Big)G^{\mathrm{R}}(\boldsymbol{r},\boldsymbol{r}_0;t,t_0)=\mathrm{i}\hbar\delta(\boldsymbol{r}-\boldsymbol{r}_0)\frac{\partial\theta(t-t_0)}{\partial t}. \tag{A13}$$

Please note that (A6) is symmetric with respect to $(t-t_0)$ and $-(t-t_0)$, and has no physical meaning of time retard or advance. Naturally, its solutions should not have the meaning of time retard or advance. Equation (A13) is valid for time $t>t_0$, such that it embodies the meaning of time retard, and so does its solution (A12).

It is seen from (A10)-(A12) that the analytic continuation endows the solution the meaning of time retard, such that it actually changes the solution. The time-retarded solution ought to be solved from a differential equation that has the meaning of time retard. That is to say, (A12) should be the solution of (A13) instead of (A6).

Lippmann and Schwinger started from Eq. (A1) and arrived at the famous equation (A5). Now, if we start from (A5) or (2.41) to do inverse Fourier transformation, we will arrive at (2.3) that is the solution of (2.2) but not of (A1).

In the present work, we start from Eq. (2.2) that contains the initial condition already, and solve it rigorously without any other assumption, so as to obtain (2.41). Reversely, Eq. (2.41) can be taken inverse Fourier transformations to verify that it satisfies Eq. (2.2). This reflects that our derivation is rigorous.

In summary, the so-call boundary condition (A3) is not an initial condition, and it is called a manual operation. The insertion of the factor $\mathrm{e}^{\varepsilon t}$ into the integral expression of the solution and the analytic continuation such as (A10) actually change the solution. These manipulations are not the kind of rigorous derivation, and we call them manual operations.

## Appendix B. The formula of general term of Eq. (2.44)

Making use of the second expression of the $G^{\mathrm{R}(0)}$ in (2.23), we are able to put down a uniform expression for any order term in (2.44).

Equation (2.44) is written as the summation of all-order terms:

$$C = C^{(0)} + C^{(1)} + C^{(2)} + C^{(3)} + \cdots, \tag{B1}$$

where the $n$th order term is

$$C^{(n)} = \int \mathrm{d}R \int \mathrm{d}R' \psi^*(R) G^{\mathrm{R}(n)}(R, R')\psi(R'). \tag{B2}$$

With the help of (2.23), we obtain the first to third terms in the following forms.

$$C^{(1)} = \int \mathrm{d}R \int_{t_0}^{t} \mathrm{d}t_1 [\mathrm{e}^{\mathrm{i}H_0(R)(t-t_0)}\psi^*(R)] H_1(R) \mathrm{e}^{-\mathrm{i}H_0(R)(t_1-t_0)}\psi(R). \tag{B3}$$

$$\begin{aligned} C^{(2)} = &\int \mathrm{d}R \int_{t_0}^{t} \mathrm{d}t_1 \mathrm{d}t_2 [\mathrm{e}^{\mathrm{i}H_0(R)(t-t_0)}\psi^*(R)] H_1(R) \\ &\mathrm{e}^{-\mathrm{i}H_0(R)(t_1-t_0)}[H_1(R)\mathrm{e}^{-\mathrm{i}H_0(R)(t_2-t_0)}\psi(R)]. \end{aligned} \tag{B4}$$

$$\begin{aligned} C^{(3)} = &\int \mathrm{d}R \int_{t_0}^{t} \mathrm{d}t_1 \mathrm{d}t_2 \mathrm{d}t_3 [\mathrm{e}^{\mathrm{i}H_0(R)(t-t_0)}\psi^*(R)] H_1(R) \\ &\mathrm{e}^{-\mathrm{i}H_0(R)(t_1-t_0)}\{H_1(R)\mathrm{e}^{-\mathrm{i}H_0(R)(t_2-t_0)}[H_1(R)\mathrm{e}^{-\mathrm{i}H_0(R)(t_3-t_0)}\psi(R)]\}. \end{aligned} \tag{B5}$$

By induction, the $n$th order term is

$$C^{(n)} = \int \mathrm{d}R \int_{t_0}^{t} \mathrm{d}t_1 \cdots \mathrm{d}t_n (A_0\psi)^* H_1 A_1 (H_1 A_2 \cdots (H_1 A_{n-1}(H_1 A_n \psi))), \tag{B6}$$

where

$$A_0 = \mathrm{e}^{-\mathrm{i}H_0(t-t_0)}; A_n = \mathrm{e}^{-\mathrm{i}H_0(t_n-t_0)}, n \ge 1. \tag{B7}$$

How to make use of this formula is still under investigation.

## References

[1] Schiff L I 1968 *Quantum Mechanics* 3rd ed. (New York: McGraw Hill Book

Company) Chap. 9
[2] Landau L D and Lifshitz E M 1977 *Quantum Mechanics Non-relativistic Theory, Vol. 3 of course of Theoretical Physics* (New York: Pergmon Press) Chap. 7
[3] Yndurain F J 1996 *Relativistic Quantum Mechanics and Introduction to Field Theory* (Berlin: Springer-Verlag) Chap. 7
[4] Wachter A 2010 *Relativistic Quantum Mechanics* (Dordrecht: Springer Science + Business Media B. V. 978-90-481-3644-5) https://doi.org/10.1007/978-90-481-3645-2
[5] Messiah A 2014 *Quantum Mechanics I* 2nd ed. (New York: Dover Publications) Chap. 10
[6] Ballentine L E 2015 *Quantum Mechanics A Modern Development* 2nd ed. Singapore: World Scientific) Chap. 16
[7] Lippmann B A and Schwinger J 1950 Variational Principles for Scattering Processes. I *Phys. Rev.* **79**(3) 469-480 https://doi.org/10.1103/PhysRev.79.469
[8] Gell-Mann M and Goldberger M L 1953 The formal theory of scattering *Phys. Rev.* **91**(2) 398-408 https://doi.org/10.1103/PhysRev.91.398
[9] Gerjuoy E 1958 Time-Independent Nonrelativistic Collision Theory *Annals of Physics* **6** 58-93 doi:10.1016/0003-4916(58)90004-6
[10] Jordan T F 1962 The Quantum-Mechanical Scattering Problem *J. Math. Phys.* **3** 414-428 https://doi.org/10.1063/1.1724242
[11] Jordan T F 1962 The Quantum Mechanical Scattering Problem. II. Multi Channel Scattering *J. Math. Phys.* **3** 429-439 doi: 10.1063/1.1724243
[12] Briggs J S and Feagin J M 2014 Scattering theory, multiparticle detection, and time *Phys. Rev. A* **90** 052712 https://doi.org/10.1103/PhysRevA.90.052712
[13] Goldberger M L 1951 Note on the general theory of scattering *Phys. Rev.* **82** 757 https://doi.org/10.1103/PhysRev.82.757
[14] Goldberger M L 1951 Approximation methods in the theory of scattering *Phys. Rev.* **84**(5) 929-938 https://doi.org/10.1103/PhysRev.84.929
[15] Corinaldesi E, Trainor L and Wu T Y 1952 The Oppenheimer approximation for the scattering of electrons, Il *Nuovo Cimento* **9**(5) 436–439 doi:10.1007/bf02783656
[16] Hack M N 1954 Bound states and the formal theory of scattering *Phys. Rev.* **96**(1) 196-198 https://doi.org/10.1103/PhysRev.96.196
[17] Lippmann B A 1956 Rearrangement Collisions *Phys. Rev.* **102**(1) 264-268 https://doi.org/10.1103/PhysRev.102.264
[18] Foldy L L and Tobosman W 1957 Application of formal scattering theory to many-body problems *Phys. Rev.* **105**(3) 1099-1100 https://doi.org/10.1103/PhysRev.105.1099
[19] Gerjuoy E 1958 Outgoing Boundary Condition in Rearrangement Collisions *Phys. Rev.* **109**(5) 1806-1814 https://doi.org/10.1103/PhysRev.109.1806
[20] Kowalski K L and Feldman D 1961 Transition Matrix for Nucleon-Nucleon Scattering *J. Math. Phys.* **2** 499-511 https://doi.org/10.1063/1.1703736
[21] Bransden B M and Moorhouse R G 1958 On the elastic scattering of pions by deuterons *Nucl. Phys.* **6** 310-319 doi: 10.1016/0029-5582(58)90111-1
[22] Levin F S 1965 Coupled-equations method for the scattering of identical particles *Phys. Rev.* **140**(4) B1099-B1109 https://doi.org/10.1103/PhysRev.145.B1099

[23] Dolinszky T 1965 On the formal theory of elastic scattering *Nucl. Phys.* **71**(2) 465-478 doi:10.1016/0029-5582(65)90734-0
[24] Rashid A M H-A and Samaranayake V K 1970 Elastic scattering of low-energy pions by alpha particles *Nucl. Phys.* B **17**(1)189-205 doi:10.1016/0550-3213(70)90409-810.1
[25] Glöckle W 1970 A new approach to the three-body problem *Nucl. Phys.* A**141** 620-630 https://doi.org/10.1016/0375-9474(70)90992-9
[26] Levin F S and Sandhas W 1984 Triad of homogegeous and inhomogeneous three-particle Lippmann-Schwinger equations *Phys. Rev. C* **29**(5) 1617-1627 https://doi.org/10.1103/PhysRevC.29.1617
[27] Benoist-Gueutal P 1986 Triad of three-particle Lippmann-Schwinger equations *Phys. Rev. C* **33**(2) 412-416 https://doi.org/10.1103/PhysRevC.33.412
[28] Gerjuoy E and Adhikari S K 1986 Alternative interpretations of the many-particle Lippmann-Schwinger equation, *Phys. Rev. C* **34**(1) 1-13 https://doi.org/10.1103/PhysRevC.34.1
[29] Gerjuoy E 1986 Comment on "Triad of three-particle Lippmann-Schwinger equations" *Phys. Rev.* C **34**(3) 1143-1148 https://doi.org/10.1103/PhysRevC.34.1143
[30] Gerjuoy E and Adhikari S K 1986 Surface integrals in the derivation of the Lippmann-Schwinger equation for many-particle systems *Phys. Lett. A* **115**(1,2) 1-5 doi:10.1016/0375-9601(86)90095-2
[31] Gerjuoy E and Adhikari S K 1987 Lippmann-Schwinger equation in a soluble three-body model: surface integrals at infinity *Phys. Rev. C* **35**(2) 415-429 https://doi.org/10.1103/PhysRevC.35.415
[32] Muga J G 1988 Invariants in potential scattering *Physica A* **153** 636-651 DOI: 10.1016/0378-4371(88)90247-6
[33] Muga J G and Levine R D 1989 Stationary scattering theories *Physica Scripta* **40** 129-140 DOI: 10.1088/0031-8949/40/2/001
[34] Gadella M and Gomez F 2002 The Lippmann–Schwinger equations in the rigged Hilbert space *J. Phys. A: Math. Gen.* **35**(40) 8505–8511 doi:10.1088/0305-4470/35/40/309
[35] Wang H Y 2022 The mathematical physical equations satisfied by the retarded and advanced Green's functions *Physics Essays* **35**(4) 380-391 http://dx.doi.org/10.4006/0836-1398-35.4.380
[36] Bohm A, Kielanowski P and Wickramasekara S 2006 Complex energies and beginnings of time suggest a theory of scattering and decay *Annals of Physics* **321** 2299–2317 doi:10.1016/j.aop.2006.02.016
[37] Wang H Y 2012 *Green's Function in Condensed Matter Physics* (Beijing: Alpha Science International ltd. and Science Press)
[38] Wang H Y 2017 *Mathematics for Physicists* (Singapore and Beijing: World Scientific Publishing Company and Science Press)
[39] Rockmore R M 1962 A Reduction Formalism for Potential Scattering *Ann. Phys.* **20**(3) 375-382 https://doi.org/10.1016/0003-4916(62)90153-7
[40] Lifshitz E M and Pitaevskii L P 1981 *Physical Kinetics, Vol. 10 of Course of Theoretical Physics* (Oxford: Pergamon Press) Chap. 1

[41] Haar D ter 1955 Foundations of Statistical Mechanics *Rev. Mod. Phys*. **27**(3) 289-338 https://doi.org/10.1103/RevModPhys.27.289
[42] Kubo R, Toda M and Hashitsume N 1985 *Statistical Physics II Nonequilibrium Statistical Mechanics* (Berlin: Springer-Verlag)
[43] Huang K 1987 *Statistical Mechanics* 2nd ed. (New York: John Wiley & Sons)
[44] Schwable F 2006 *Statistical Mechanics* 2nd ed. (Berlin: Springer-Verlag)
[45] Bjorken J D and Drell S S 1964 *Relativistic quantum mechanics* (New York: McGraw-Hill Book Co.)
[46] Abers E S 2004 *Quantum Mechanics* (Upper Saddle River, New Jersey: Pearson Education Inc.)
[47] Kleinert H 2009 *Path Integrals in Quantum Mechanics, Statistics, Polymer Physics, and Financial Markets* 5th ed. (Singapore: World Scientific Publishing Co. Pte. Ltd.) Chap. 1
[48] Greiner W and Reinhardt J 2009 *Quantum Electrodynamics* 4th ed. (Berlin, Heidelberg: Springer-Verlag)
[49] Hibbs A R and Feynman R P 1965 *Quantum Mechanics and Path Integrals* (New York: McGraw-Hill, Inc.)
[50] Scheck F 2007 *Quantum Physics* (Berlin, Heidelberg: Springer-Verlag)
[51] Mahan G H 2009 *Quantum Mechanics in a Nutshell* (Princeton and Oxford: Princeton University Press)
[52] Bransden B H and Joachain C J 2000 *Quantum Mechanics* 2nd ed. (London: Pearson Education Limited)
[53] Bes D R 2007 *Quantum Mechanics A Modern and Concise Introductory Course* 2nd ed. (Berlin, Heidelberg: Springer-Verlag)
[54] Wang H Y 2022 Evaluation of cross section of elastic scattering for non-relativistic and relativistic particles by means of fundamental scattering formulas *Advanced Studies in Theoretical Physics* **16**(3) 131-163
https://doi.org/10.12988/astp.2022.91866
[55] Wang H Y 2023 Liouville equation in statistical mechanics is not applicable to gases composed of colliding molecules *Physics Essays* **36**(1) 429-437
http://dx.doi.org/10.4006/0836-1398-36.1.429
[56] Tien C L and Lienhard J H 1979 *Statistical thermodynamics* (New York: Hemisphere Pub. Corp., Washingtong; McGraw-Hill)
[57] Liboff R L 2003 *Introduction to the theory of kinetic equations* (New York: Robert E. Krieger Publishing Company, Huntington)
[58] Liboff R L 2003 *Kinetic theory: classical, quantum, and relativistic descriptions* 3rd ed. (New York: Springer-Verlag)
[59] Bellac M Le, Mortessagne F and Batrouni G G 2004 *Equilibrium and non-equilibrium statistical thermodynamics* (Cambridge: Cambridge University Press)
[60] Oppenheimer J R 1928 *Phys. Rev*. **32** 361-376
https://doi.org/10.1103/PhysRev.32.361
[61] Wang H Y 2021 Fundamental formalism of statistical mechanics and thermodynamics of negative kinetic energy systems *J. Phys. Commun*. **5** 055012
https://doi.org/10.1088/2399-6528/abfe71

[62] Wang H Y 2020 New results by low momentum approximation from relativistic quantum mechanics equations and suggestion of experiments *J. Phys. Commun.* **4** 125004 https://dx.doi.org/10.1088/2399-6528/abd00b
[63] Wang H Y 2022 The modified fundamental equations of quantum mechanics *Physics Essays* **35**(2) 152-164 http://dx.doi.org/10.4006/0836-1398-35.2.152
[64] Wang H Y 2020 Solving Klein's paradox *J. Phys. Commun.* **4** 125010 https://doi.org/10.1088/2399-6528/abd340
[65] Wang H Y 2021 Macromechanics and two-body problems *J. Phys. Commun.* **5** 055018 https://doi.org/10.1088/2399-6528/ac016b
[66] Wang H Y 2021 Virial theorem and its symmetry *J. of North China Institute of Science and Technology* **18**(4) 1-10 (in Chinese) http://10.19956/j.cnki.ncist.2021.04.001
[67] Wang H Y 2022 Resolving problems of one-dimensional potential barriers based on Dirac equation *J. of North China Institute of Science and Technology* **19**(1) 97-107 (in Chinese) http://10.19956/j.cnki.ncist.2022.01.016
[68] Wang H Y 2022 There is no vacuum zero-point energy in our universe for massive particles within the scope of relativistic quantum mechanics *Physics Essays* **35**(3) 270-275 http://dx.doi.org/10.4006/0836-1398-35.3.270
[69] Wang H Y 2023 The behaviors of the wave functions of small molecules with negative kinetic energies *Physics Essays* **36**(2) 140-148 http://dx.doi.org/10.4006/0836-1398-36.2.140
[70] Wang H Y 2023 A theory of dark energy that matches dark matter *Physics Essays* **36**(2) 149-159 http://dx.doi.org/10.4006/0836-1398-36.2.149
[71] Wang H Y 2023 Many-Body Theories of Negative Kinetic Energy Systems *Physics Essays* **36**(2) 198-211 http://dx.doi.org/10.4006/0836-1398-36.2.198"